\documentclass[11pt]{article}
\usepackage[hmargin=1in,vmargin=1in]{geometry}
\usepackage{hchang}
\usepackage{thm-restate}
\usepackage{cleveref}
\usepackage{comment}
\usepackage{tcolorbox}
\usepackage{soul}

\nolinenumbers

\newtheorem{theorem}{Theorem}[section]
\newtheorem{lemma}[theorem]{Lemma}
\newtheorem{claim}[theorem]{Claim}
\newtheorem{observation}[theorem]{Observation}

\newtheorem{remark}[theorem]{Remark}
\newtheorem{corollary}[theorem]{Corollary}

\newtheorem{definition}[theorem]{Definition}

\newcommand{\eps}{\varepsilon}

\newcommand{\cP}{\mathcal{P}}
\newcommand{\cT}{\mathcal{T}}
\newcommand{\cR}{\mathcal{R}}

\newcommand{\cX}{\mathcal{X}}
\newcommand{\dist}{\ensuremath{\delta}}

\newcommand{\spath}{\operatorname{path}}
\newcommand{\safe}{\operatorname{Safe}}

\newcommand{\proxy}{\operatorname{Proxy}}
\newcommand{\relP}{\operatorname{RelPairs}}

\newcommand{\domain}{\operatorname{dom}}

\newcommand{\replace}[2]{\ensuremath{\operatorname{Swap}(#1, #2)}}
\newcommand{\domreplace}[3]{\ensuremath{\operatorname{Swap}_{#1}(#2, #3)}}
\newcommand{\bdry}{\partial\!}
\newcommand{\spread}{D}

\newcommand{\canon}[2]{\ensuremath{#1{\rightarrow}#2}}

\title{Distance Approximating Minors for Planar and Minor-Free Graphs%
\thanks{Hsien-Chih Chang and Jonathan Conroy are supported by the U.S.\ National Science Foundation CAREER Award under the Grant No.\ CCF-2443017.}}
\author{%
Hsien-Chih Chang%
\thanks{Department of Computer Science, Dartmouth College. Email: {\tt hsien-chih.chang@dartmouth.edu}.} 
\and 
Jonathan Conroy%
\thanks{Department of Computer Science, Dartmouth College. Email: {\tt jonathan.conroy.gr@dartmouth.edu}}  
}
\date{}

\begin{document}

\maketitle
\thispagestyle{empty}

\begin{abstract}
Given an edge-weighted graph $G$ and a subset of vertices $T$ called terminals, an $\alpha$-distance-approximating minor ($\alpha$-DAM) of $G$ is a graph minor $H$ of $G$ that contains all terminals, such that the distance between every pair of terminals is preserved up to a factor of $\alpha$.
Distance-approximating minor would be an effective distance-sketching structure on minor-closed family of graphs; in the constant-stretch regime it generalizes the well-known Steiner Point Removal problem by allowing the existence of (a small number of) non-terminal vertices.
Unfortunately, in the $(1+\varepsilon)$ regime the only known DAM construction for planar graphs relies on overlaying $\tilde{O}_\varepsilon(|T|)$ shortest paths in $G$, which naturally leads to a quadratic bound in the number of terminals [Cheung, Goranci, and Henzinger, ICALP 2016].

We break the quadratic barrier and build the first $(1+\varepsilon)$-distance-approximating minor for $k$-terminal planar graphs and minor-free graphs of near-linear size
$\tilde{O}_\varepsilon(k)$.
In addition to the near-optimality in size, the construction relies only on the existence of shortest-path separators [Abraham and Gavoille, PODC 2006] and $\varepsilon$-covers [Thorup, J.\ ACM 2004]. 
Consequently, this provides an alternative and simpler construction to the near-linear-size emulator for planar graphs [Chang, Krauthgamer, and Tan, STOC 2022], as well as the first near-linear-size emulator for minor-free graphs. Our DAM can be constructed in near-linear time.
\end{abstract}
\newpage

\thispagestyle{empty}
\tableofcontents

\newpage
\setcounter{page}{1}

\section{Introduction}

Distance compression and sketching have proven to be ubiquitous in modern graph algorithm design.
In this paper we study a natural distance sketching structure on planar and minor-free graphs whose existence is unknown but has been alluded to, with algorithmic consequences~\cite{cgh+-fdcde-2020a,ckt-lpg-2022}.
Let $G$ be an undirected edge-weighted graph from some minor-closed family.
Our goal is to preserve pairwise distances between a subset of vertices in $G$ called \EMPH{terminals}; we denote the set of terminals as $T$.
A \EMPH{stretch-$\alpha$ distance-approximating minor} of $G$ (shorthanded as \EMPH{$\alpha$-DAM})~\cite{knz-ptdum-2014,CGH16} is an (edge-weighted\footnote{The weights on the edges of $H$ may be arbitrary, not necessarily tied to the graph $G$.}) graph minor $H$ of $G$ that contains all terminals in $T$ as vertices and possibly other vertices in $G$, such that for every pair of terminals $x$ and $y$,
\[
d_G(x,y) \le d_H(x,y) \le \alpha \cdot d_G(x,y).
\]
Distance-approximating minors generalize the well-studied notion of \emph{spanners}, which are required to be subgraphs of $G$.
Generally speaking, spanners cannot have size depending only on the terminal set (for example, think of a single path with only the two endpoints as terminals) and are usually measured in terms of lightness (the ratio of its weight to that of the minimum spanning tree).
Therefore spanners are useful for applications related to connectivity problems like Steiner tree and TSP~\cite{kle-sspga-2006,bkm-lasst-2009}, but unable to help with distance-related optimization tasks like distance oracles or (approximate) diameter.
On the other hand, distance-approximating minors retain the structure of the input graph $G$ compared to more relaxed distance sketching structures like \emph{emulators}, where the vertices and edges can come from out of nowhere (unrelated to $G$) as long as the terminals in $T$ are still present and their distances are preserved pairwise.
Emulators can be useful for distance inquiries on planar graphs (after submitting to the standard graph decomposition pipeline using $r$-division~\cite{fre-faspp-1987,kms-srsdp-2013}), but fall short in their ability to return an actual path realizing the approximate distance.

Constructing distance-approximating minors has proven to be difficult, even for popular minor-closed families like planar graphs.
It is straightforward to construct (exact) DAM in any graph with $O(|T|^4)$ vertices by overlaying the (unique) shortest path between every pair of terminals in $G$ \cite{knz-ptdum-2014}; however it is impossible to construct an exact DAM with subquadratic size~\cite{knz-ptdum-2014,co-pemm-2020}, even when the input graph is a planar grid. 
At the other extreme, if we restrict the size of the DAM to be exactly $|T|$, forbidding any non-terminal vertex to occur in $H$, then this is the famous \emph{Steiner Point Removal problem}~\cite{Gupta01,BG08,kkn-ccchr-2015,che-sprdt-2018,Filtser18,FKT19,Filtser20,HL22}, where the lower bound on stretch is at least $8(1-o(1))$~\cite{CXKR06} when the graph is minor-free, and
getting $O(1)$-stretch was an open problem for two decades until recently being resolved by Chang, Conroy, Le, Milenković, Solomon, and Than~\cite{Filtser20B,ccl+-cpmot-2023,ccl+-spmgs-2024} on planar and minor-free graphs.

In this paper we focus on the $(1+\e)$ stretch regime.
In this regime, we incur extremely small stretch while retaining hope to bypass the $\Omega(|T|^2)$ lower bound from exact DAM; indeed, the planar-grid lower bound construction in the exact setting only yields an $\Omega(|T|/\eps)$ lower bound in the $(1+\e)$-approximate setting~\cite{knz-ptdum-2014}. As we will describe later, achieving subquadratic size would be essential for several algorithmic applications.
The best construction for planar graphs to date has size $\tilde{O}(|T|^2/\eps^2)$\footnote{Here, the $\tilde O(\cdot)$ notation hides factors of $\poly (\log |T|)$. Later in the paper, we consider graphs with $n$ vertices and edge weights between $1$ and $\Phi$, and we use the notation $\tilde O(\cdot)$ to hide factors of $\poly (\log n, \log \Phi)$; see \Cref{rem:spread}.}, by Cheung, Goranci, and Henzinger \cite{CGH16}. They define $\tilde{O}(|T|/\eps)$ shortest paths based on \emph{path separators} and \emph{$\e$-covers}~\cite{Thorup04}, and \emph{overlay} the shortest paths by taking the union; as any pair of shortest paths could intersect, this leads to a size of $\tilde{O}(|T|^2/\eps^2)$.
(We observe that the same construction can be generalized to minor-free graphs if we plug in the path separators for minor-free graphs~\cite{ag-olups-2006}.)
%
The quadratic bound in DAM size seems to be an unavoidable barrier for the approach of overlaying shortest paths: we may need at least $\Omega(|T|)$ shortest paths to even connect all the terminals, and simply overlaying them leads to $\Omega(|T|^2)$ size.
The natural open question is: 
\begin{quote}
\emph{Does every minor-free graph family have $(1+\e)$-DAMs of size near-linear in $|T|$, or does a quadratic lower bound exist?}
\end{quote}
%

A recent breakthrough result by Chang, Krauthgamer, and Tan~\cite{ckt-lpg-2022,ckt-aepg-2022}
reignited the search for a near-linear-size $(1+\e)$-DAM.  
Specifically, they showed that an \emph{emulator} of stretch $1+\e$ and size $\tilde O_\e (|T|)$ exists for any planar graph $G$; moreover, the emulator $H$ \emph{itself} is a planar graph (although not a minor of $G$).
This leads to various applications including ultra-efficient computation of (approximate) shortest-path, minimum cut, diameter, and offline dynamic distance oracle.
Most recently, a \emph{fully dynamic} distance oracle for planar graphs with $n^{o(1)}$ query and update time was built upon the CKT emulator~\cite{fgpp-ovfa-2024}.
Unfortunately, the CKT construction relies heavily on planarity, specifically the planar embedding.  They performed a \emph{split-and-combine} strategy by cutting and slicing the planar graph into pieces each having exactly one boundary component in the planar embedding (which can only be achieved in planar graphs), and then proceed to identify a carefully chosen shortest-path separator that splits the piece in a (roughly) balanced fashion and compute $\e$-cover on the path separator.
While shortest-path separators exist for minor-free graphs~\cite{ag-olups-2006}, the choice of the shortest-path separator here is very delicate and crucially relies on the piece being planar with exactly one boundary to work.  
As a result, there is no way to mimic the CKT construction for general minor-free graphs
(or even for apex-minor-free graphs).

\subsection{Main Results}

The main result of this paper is the following theorem.  (For a precise statement, see Theorem~\ref{thm:dam-scaled}.)

\begin{theorem}
\label{thm:dam}
Let $G$ be an $n$-vertex minor-free graph with edge weights between $1$ and $\Phi$, let $T$ be a set of terminals $T$ in $G$, and let $\e$ be a parameter in $(0,1)$. 
There is a stretch $(1+\e)$ distance-approximating minor for $G$ with respect to terminals in $T$, with size $|T| \cdot \poly(\log n, \log \Phi, \e^{-1})$.
\end{theorem}

Our result gives the first ever construction for $(1+\e)$-DAM on planar graphs, as well as on minor-free graphs, that has size near-linear in $|T|$.
From a technical point of view, our strategy is (perhaps surprisingly) a refinement of the shortest-path overlay approach.
Our construction bypasses the need for a planar embedding and directly works with shortest-path separators and $\e$-covers; not only does this makes the construction generalizable to minor-free graphs, but also provides a new alternative construction to the $(1+\e)$-emulator for planar graphs that are cleaner both in description and in proofs. 
(The CKT construction spans 50 pages and is quite delicate.)
To avoid the quadratic blowup from overlaying shortest paths, we carefully \emph{detour} the paths to construct a set of \emph{approximate} shortest paths that (in some amortized sense) intersect sparsely with each other.
We believe this technique may be of independent interest.

\begin{remark}
\label{rem:spread}
One drawback of our \Cref{thm:dam} is that the size of DAM has a poly log dependence on the graph size $n$ and the maximum edge weight $\Phi$, whereas previous results (namely, the DAMs of \cite{knz-ptdum-2014} and \cite{CGH16}, and the emulator of \cite{ckt-lpg-2022}) depend only on the number of terminals $|T|$. We could remove the dependence on $n$ by first computing the exact $O(|T|^4)$-size DAM of \cite{knz-ptdum-2014} and then applying our DAM on top of the $O(|T|^4)$-size graph, but it is unclear how to remove the dependence on $\Phi$. These extra dependencies don't bother much for our applications, but it would be interesting to construct a DAM without dependence on $\Phi$.
\end{remark}

\paragraph{Applications.}
In minor-free graphs, DAM is particularly helpful because it enables efficient distance compression on regions of an \EMPH{$r$-division} of the input graph and can thus lead to better divide-and-conquer algorithms. As an illustrative example, we describe how to use $r$-division to \emph{bootstrap} and obtain a linear-time construction of our DAM.
This is somewhat surprising because our DAM construction relies on computing path separators in minor-free graphs, and the best running time so far remains quadratic~\cite{ag-olups-2006,KKR12}. 
(We did not improve the construction time for path separators; it remains to be an interesting open question.)

\begin{restatable}{theorem}{thmFastFull}
\label{thm:fast-full-statement}
Given an $n$-vertex minor-free graph $G$ with edge weights between $1$ and $\Phi$, a set of terminals $T$ in $G$, and a parameter $\e$ in $(0,1)$, one can compute a $(1+\e)$-DAM of size $|T| \cdot \poly(\log n, \log \Phi, \e^{-1})$ in near-linear time $n \cdot \poly(\log n, \log \Phi, \e^{-1})$.
\end{restatable}
We prove \Cref{thm:fast-full-statement} formally in Section~\ref{S:fast}, but here we sketch the main idea to illustrate the power of DAM, inspired by the fast construction of \cite{ckt-lpg-2022}. An \EMPH{$(r,s)$-division} of a graph $G$ is a partition of the edges of $G$ into $O(n/r)$ \emph{regions}, such that each region $R$ (1) has at most $r$ vertices and (2) shares at most $O(s)$ \emph{boundary vertices} with other regions. 
An $(r,r^{1-\delta})$-division for $\delta \in (0,1)$ is called an \EMPH{$r$-division}.
While $(r,\sqrt{r})$-divisions are known to exist in any minor-free graphs~\cite{AST90}, for our purpose we will use 
an $(r, r^{2/3} \log^{1/3} r)$-division that can be constructed in $O(n)$ time~\cite{HR24}.
First we compute an $r$-division for $G$ by setting $r \coloneqq (\log n)^{O(1)}$, then apply the $(1+\e)$-DAM construction on each region $R$ for terminals in $T \cap R$ \emph{and} all boundary vertices of $P$.
By gluing the region-wise DAMs together we obtain a global $(1+\e)$-DAM for the original graph $G$, whose size is at least a factor $(\log n)^{O(1)}$ smaller than $G$ --- intuitively, each piece has $r$ vertices but only $\tilde O(r^{2/3})$ boundary vertices, so replacing the piece with our near-linear-size DAM shrinks the size of each piece roughly by a factor $\tilde O(r^{1/3})$.
The time taken is $O(n + r^{O(1)} \cdot (n/r)) = O(n\log^{O(1)} n)$. Repeating this process $O(\log n)$ times constructs a DAM of near-linear size, in near-linear time.
Notice that this bootstrapping trick is made possible precisely because we are computing a distance-approximating \emph{minor}; for general emulators in minor-free graphs, the glued-back global emulator may no longer be minor-free, and therefore we cannot bootstrap further.

As a corollary of our near-linear-time construction of DAM, we construct offline $(1+\e)$-approximate dynamic distance oracles via the framework of Chen \etal~\cite{cgh+-fdcde-2020a}. In this problem, we are given a graph $G$ and a sequence of \emph{updates} to $G$ (either edge insertions or edge deletions) and \emph{queries} (a pair of vertices $s$ and $t$ whose distance we must report in the current graph). The goal is to minimize the query and update time.
We consider the problem under the assumption that, at every time step, the graph excludes a fixed-size minor and has edge weights between 1 and $\poly(n)$.
Chen \etal~\cite{cgh+-fdcde-2020a} show that, in this setting, an offline $(1+\e)$-approximate dynamic distance oracle can be constructed if every minor-free graph with edge weights between $1$ and $\Phi$\footnote{The reduction of \cite{cgh+-fdcde-2020a} isn't phrased with the restriction on edge weights, but it is immediate that their proof applies also to this setting.} has a $(1+\e)$-DAM of size $|T| \cdot \poly(\log n, \log \Phi, \e^{-1})$ that can be constructed in time $n \cdot \poly(\log n, \log \Phi, \e^{-1})$.

\begin{theorem}
There is an offline dynamic $(1+\e)$-approximate distance oracle with $\poly(\log n, \log \Phi, \e^{-1})$ query and update time, for $n$-vertex minor-free graphs with edge weights between $1$ and $\Phi$.
\end{theorem}

\subsection{Technical Overview}
\label{SS:overview}
Let $G$ be a connected $n$-vertex planar graph with diameter $D$, and let $T$ be a set of terminals. 
We use the notation $\tilde O_\e(\cdot)$ to hide factors of $\poly(\log n, \log D, \e^{-1})$.

\paragraph{Review: A \boldmath{$(1+\e)$} non-planar emulator.}
We first review a simple construction of a near-linear size $(1+\e)$-stretch \emph{non-planar} emulator, which is mostly implicit in the work of \cite{Thorup04} and \cite{CGH16} (see also a comment on Stack Exchange by Christian Sommer \cite{sommerStackExchange}).
We go through the result slowly so that we can introduce all the terminology we will use for our later construction of DAM.
We begin with the classic shortest-path separator theorem of Lipton and Tarjan \cite{lt-stpg-1979,GKR01,Thorup04}: in any $n$-vertex planar graph, there is a set of $O(1)$ shortest paths whose removal splits the graph into connected subgraphs (which we often call \EMPH{regions}) each with at most $2n/3$ vertices. Recursively applying the shortest-path separator theorem on the resulting pieces yields a \EMPH{separator hierarchy}: a hierarchical partition of the graph into regions. For each region $R$ in the separator hierarchy, the \EMPH{internal separator} is a set of $O(1)$ paths whose removal splits $R$ into roughly equal-sized pieces; the \EMPH{external separators} are the set of $O(\log n)$ paths which are the internal separators of all ancestor pieces of $R$. Note that any path in $G$ that starts inside region $R$ and then leaves $R$ must pass through some external separator of $R$. 

Along each separator path we place \emph{portals}, as in \cite{Thorup04}. For each region $R$, and each internal separator path $P$ for region $R$, we define the \EMPH{scale-$i$ portals} to be a minimal set of vertices on $P$ such that every vertex of $P$ is within distance $\le \e 2^i$ of some scale-$i$ portal.%
\footnote{For the sake of simplicity, in this section we assume the diameter $D$ is $\poly(n)$, so we only need to consider $O(\log n)$ scales.}
Notice that for every vertex $v \in R$, there are only $O(\e^{-1})$ scale-$i$ portals on $P$ \emph{that are within distance $O(2^i)$ of $v$}. If $p$ is a scale-$i$ portal within distance $O(2^i)$ of vertex $v$, we say that $p$ is a \EMPH{relevant portal} for $v$. Intuitively, the relevant portals of $v$ are enough to capture $(1+\e)$-approximate distances between $v$ and any vertex on $P$.%
\footnote{In fact, \cite{Thorup04} gave a different argument involving a potential function that shows that the distances between $v$ and $P$ can be captured with only $O(\e^{-1})$ portals across all scales, as opposed to $O(\e^{-1} \log n)$ portals as we have described here. 
For technical reasons, the construction of the non-planar emulator using scale-$i$ portals is simpler than the construction using \cite{Thorup04} portals, and later (in the construction of our DAM) it becomes essential to use scale-$i$ portals; this is because the scale-$i$ portals are defined independently of the terminals and can be ``shared'' by nearby terminals, whereas \cite{Thorup04} would choose different portals for every terminal.} 
We construct our non-planar $(1+\e)$-distortion emulator $H$ similarly to the $+\e D$ case:
\begin{quote}
    Initialize $H$ as an empty graph with vertex set $V(G)$. For every region $R$ in the separator hierarchy, and for every terminal $t$ in $R$: add (to $H$) an edge between $t$ and every \emph{relevant portal} of $t$ on the internal separator paths of $R$.
\end{quote}
The size of $H$ is $O(|T| \e^{-1} \log^2 n)$: every terminal is contained in $O(\log n)$ regions, each of which contains $O(\e^{-1} \log n)$ relevant portals.
Moreover, one can show that $H$ preserves distances up to a $1+\e$ multiplicative factor. Indeed, let $t_1$ and $t_2$ be two terminals. Let $\pi$ be the shortest path in $G$ between $t_1$ and $t_2$, let $i$ be the smallest scale with $\dist_{G}(t_1,t_2) \le 2^i$, and let $R$ be the lowest region that contains both $t_1$ and $t_2$. Any path between $t_1$ and $t_2$ must pass through an internal or external separator of $R$; in particular, $\pi$ passes through an internal separator of some region $R'$ which is an ancestor of both $t_1$ and $t_2$. By definition of portal, there is some scale-$i$ portal $p$ (relevant to both $t_1$ and $t_2$) on an internal separator of $R'$ such that $p$ is within distance $\e \cdot 2^i$ of $\pi$. Triangle inequality implies that, in the emulator $H$, the edges $(t_1, p)$ and $(p, t_2)$ form a 2-hop path with length at most $\dist_G(t_1, t_2) + 2\e \cdot 2^i = (1+O(\e)) \cdot \dist_G(t_1,t_2)$.

\medskip The emulator $H$ is only an emulator, not a DAM. As shown in \cite{CGH16}, one can get a DAM by \emph{overlaying shortest paths}---that is, instead of creating an edge in $H$ between each terminal $t$ and portal $p$, we could add to $H$ the entire shortest path between $t$ and $p$ in $G$. After overlaying paths in this manner, we contract away all degree-2 vertices; the only remaining vertices are either terminals or vertices with degree larger than 2 (which we call \EMPH{splitting points}). Unfortunately, overlaying paths causes the size to blow up to quadratic $\tilde{O}_\e(|T|^2)$: there are $\tilde{O}_\e(|T|)$ paths that connect terminals to portals, and each pair of paths could intersect (creating up to 2 splitting points that cannot be contracted away).
For this reason, \cite{ckt-aepg-2022} required completely different techniques to achieve their near-linear size emulator. We show that, perhaps surprisingly, one can get a near-linear size DAM by overlaying an appropriate set of shortest paths. We choose these paths carefully (detouring them in a delicate way that causes them to wander in and out of each region) to bound the number of intersections.

\begin{figure}
    \centering
    \includegraphics[width=0.45\linewidth]{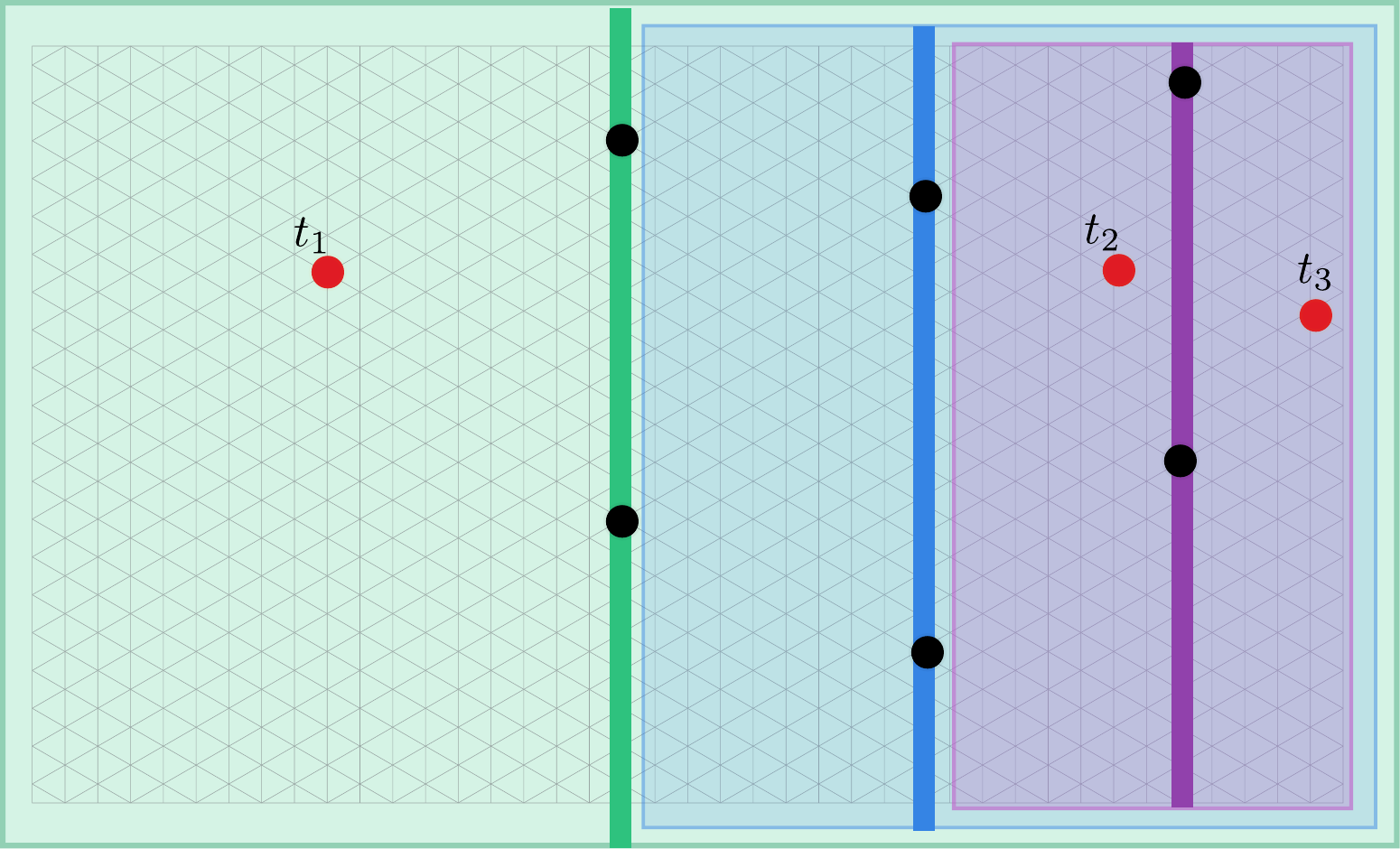}
    \hspace{2em}
\includegraphics[width=0.45\linewidth]{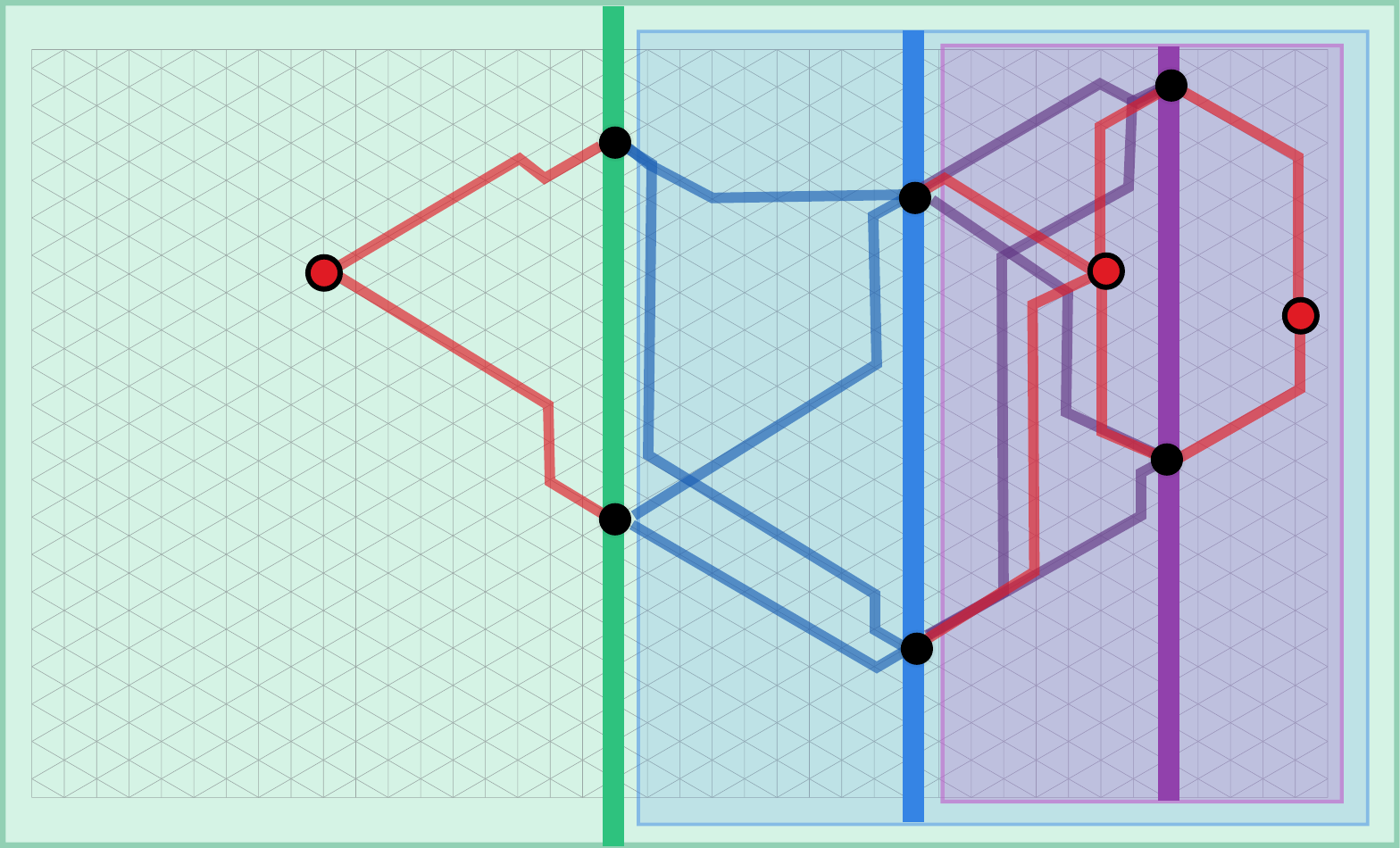}
    \caption{
 On the left: A graph $G$, three terminals $t_1$, $t_2$, $t_3$, and shaded regions representing the regions $\cR$ which contain more than one terminal. (The regions containing \emph{exactly} one terminal are not drawn.)  The internal separator of each depicted region is drawn as a solid line.
    On the right: The paths between canonical pairs and terminals that are added to the $+\e D$ DAM. For every region $R \in \cR$ which contains more than one terminal, the paths added when $R$ is processed are drawn in the same color is $R$. The paths added by regions containing exactly 1 terminal are drawn in red. Some red paths intersect (purple) paths which were added when an ancestor region was processed.}
    \label{fig:additive-dam}
\end{figure}

\paragraph{A DAM with additive \boldmath{$+\e D$} distortion.} 
As a warm-up, we construct a near-linear size DAM $H$ with additive distortion $+\e D$, for planar graphs. This guides the intuition for the $(1+\e)$ DAM. For the $+\e D$ DAM, we only use scale-$O(\log D)$ portals. As every shortest path has length at most $D$, there are only $O(\e^{-1})$ such portals per separator path.%
\footnote{Here we assume that the separator paths are shortest paths \emph{with respect to $G$}, which has diameter $D$. This assumption can be satisfied in planar graphs, but not minor-free graphs. For minor-free graphs the argument is more technical.} 
For each region $R$, we define the \EMPH{canonical pairs} of $R$ to be the pairs of scale-$O(\log D)$ portals $(p, p')$, where $p$ is a portal on some internal separator of $R$ and $p'$ is a portal on some external separator of $R$.
We construct the DAM as follows (see also \Cref{fig:additive-dam} for an illustration):
\begin{quote}
    Initialize $H$ as an empty graph with vertex set $V(G)$. Let $\mathcal{R}$ be the set of regions in the separator hierarchy which contain some terminal. For every region $R \in \mathcal{R}$: if $R$ contains more than one terminal, then for every canonical pair $(p, p')$ of $R$, add (to $H$) a shortest path in $R$%
    \footnote{Technically, the region $R$ does not include the external separators, so $p'$ is not in $R$; we must actually consider the shortest path in the graph $R^{\arcto}$ which is the subgraph induced by the union of $R$ and one of its external separators. We ignore this detail in the technical overview.} between $p$ and $p'$; alternatively,  if $R$ contains exactly one terminal $t$, then for every portal $p'$ on an external separator of $R$, add a shortest path in $R$ between $t$ and $p$. After overlaying all paths, contract away degree-2 vertices.
\end{quote}

One can show that $H$ has additive distortion $+O(\log n) \cdot \e D$ (and so rescaling $\e$ to $\e /O(\log n)$ gives the desired result). Intuitively, a path between $t_1$ and $t_2$ in $G$ can be broken up between $O(\log n)$ regions of the separator hierarchy, and we can jump between these regions with only $+\e D$ distortion by using a canonical pair. (See \Cref{S:rel-pairs} for a similar argument, in the $1+\e$ setting).

As for the size bound, we show that $H$ has size at most $O(|T|  \e^{-1}\log^2 n) = \tilde O_\e(|T|)$ via a charging argument. Imagine that we process the regions $R \in \mathcal{R}$ in a \EMPH{root-to-leaf manner}, according to the separator hierarchy. Each time we process a region $R$, we add $O(\e^{-2} \log n)$ paths to our DAM. Indeed, if $R$ contains more than one terminal, then we add at most  $O(\e^{-2} \log n)$ paths because there are only $O(\e^{-2} \log n)$ canonical pairs of $R$ (as there are $O(\log n)$ external separator paths and $O(\e^{-1})$ portals on each); if $R$ contains exactly one terminal, then we add at most $O(\e^{-1} \log n)$ paths.
We now argue that, at the time some path $\pi$ is added to $H$, it intersects only $O(\e^{-2} \log^2 n)$ paths that were added earlier. (This would prove that $H$ has size $\tilde O_\e(|T|)$, as desired.)
Indeed, for any region $R$, there are only $O(\e^{-2} \log n)$ canonical pairs of $R$, so $\pi$ only intersects $O(\e^{-2} \log n)$ other paths that were also added while region $R$ 
was processed (i.e. paths added at the same time as $\pi$). The path $\pi$ may also intersect paths that were added earlier, when some \emph{ancestor} region of (containing $R$) was processed. But there are only $O(\log n)$ such ancestor regions, and each one has only $O(\e^{-2} \log n)$ canonical pairs. This proves the bound.

\begin{figure}[h!]
    \centering
    \includegraphics[width=0.45\linewidth]{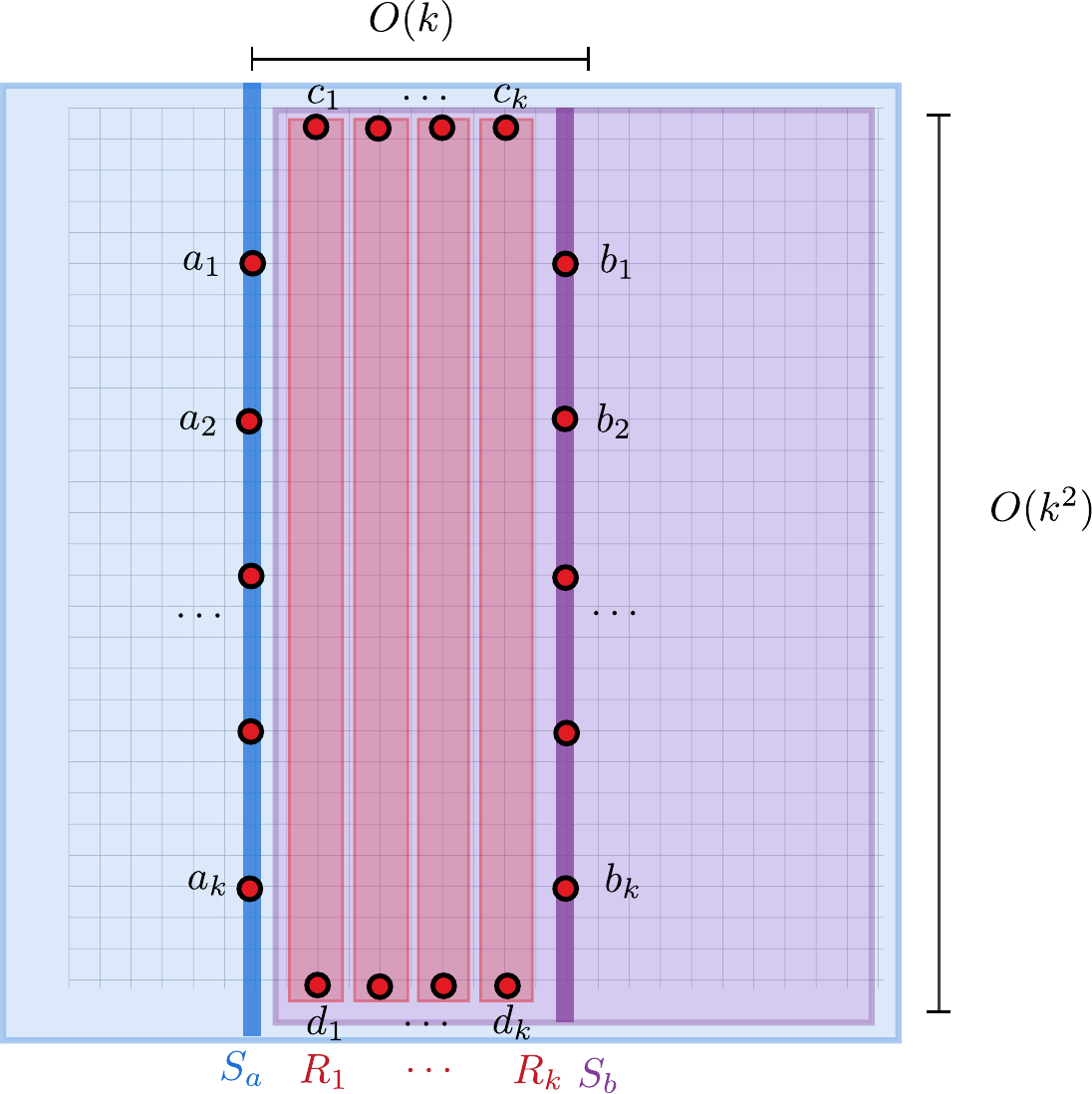}
    \hspace{3em}
    \includegraphics[width=0.37\linewidth]{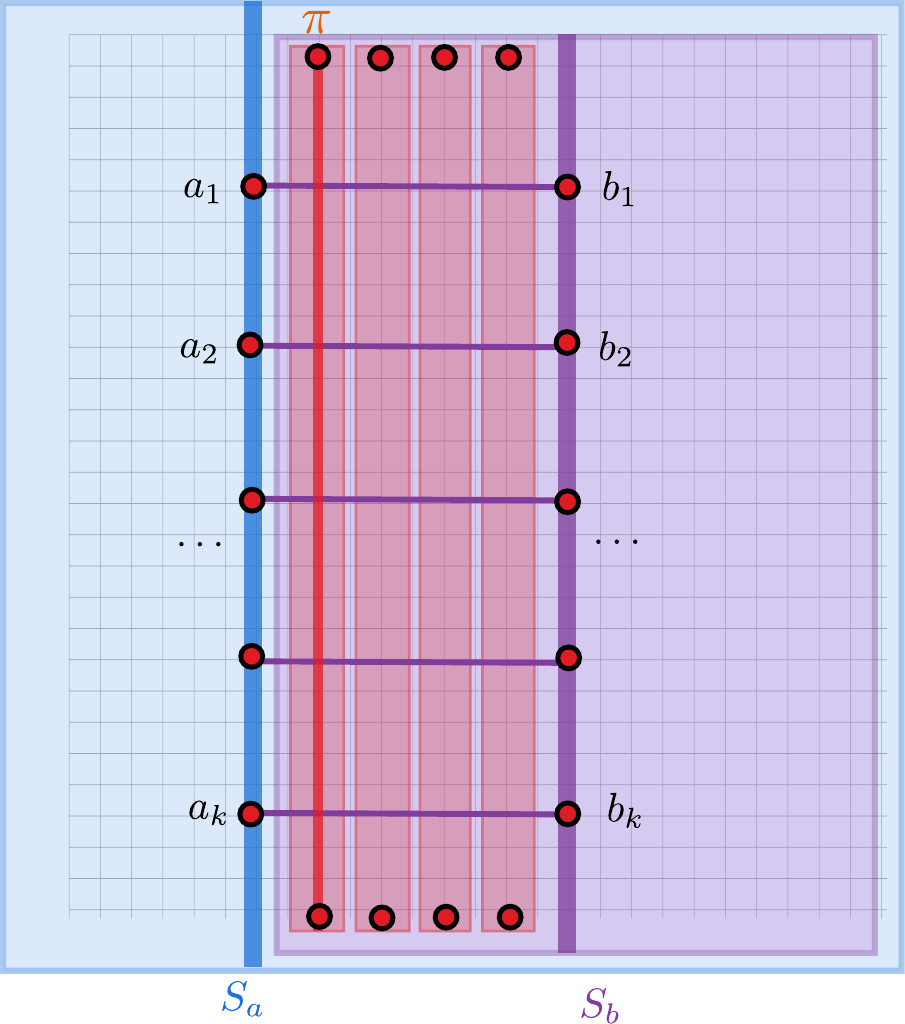}
    \caption{A rectangular $O(k) \times O(k^2)$ grid graph, and a set $T$ of $O(k)$ terminals, for which our first attempt at $(1+\e)$ DAM could have $O(|T|^2) = O(k^2)$ size.
\textbf{On the left}: A depiction of some regions in the separator hierarchy. The root region and its internal separator $S_a$ is drawn in blue. One of its children regions, and its internal separator $S_b$, is drawn in purple. The next few levels of the separator hierarchy are not drawn, but the regions $R_1, \ldots, R_k$ at the leaf level of the separator hierarchy are drawn in red; each $R_i$ consists of a single internal separator path (not drawn). The terminals are drawn in red: there are $k$ terminals on $S_a$, $k$ terminals on $S_b$, and 2 terminals per region $R_i$.
\textbf{On the right}: A partially drawn $(1+\e)$ DAM. The terminals $a_i$ and $b_i$ are themselves portals; in particular every pair $(b_i, a_i)$ is a canonical pair, and a path was added between $(b_i, a_i)$ when the purple region was processed. The terminals in the regions $R_i$ are themselves portals; in particular the pair $(c_i, d_i)$ is a canonical pair, and a path between them is added when $R_i$ is processed. The path $\pi$ between $c_1$ and $d_1$ intersects $\Omega(k)$ paths that are shorter than $\pi$.}
    \label{fig:bad-grid}
\end{figure}

\paragraph{First attempt at generalizing to \boldmath{$1+\e$} distortion.}
Let's try to generalize this $+\e D$ DAM construction to $(1+\e)$ multiplicative distortion. For each region $R$ and each scale $i$, we define the \EMPH{scale-$i$ canonical pairs} of $R$ to be the pairs of scale-$i$ portals $(p, p')$, where $p$ is a scale-$i$ portal on an internal separator of $R$, $p'$ is a scale-$i$ portal on an internal or external separator of $R$, and $\dist_R(p, p') \le O(2^i)$. For any vertex $v \in R$, the \EMPH{relevant} scale-$i$ canonical pairs are those pairs $(p,p')$ where $p$ and $p'$ are relevant scale-$i$ portals for $v$ (that is, $p$ and $p'$ are scale-$i$ portals within distance $O(2^i)$ of $v$ in $R$). As any vertex $v$ has at most $\tilde O_\e(1)$ relevant portals in $R$, it has at most $\tilde O_\e(1)$ relevant canonical pairs. Our first attempt at a $(1+\e)$ DAM is the following:
\begin{quote}
    Initialize $H$ as an empty graph with vertex set $V(G)$. Let $\mathcal{R}$ be the set of regions in the separator hierarchy which contain some terminal. For every region $R \in \mathcal{R}$, for every canonical pair $(p, p')$ of $R$ \emph{which is relevant to some terminal $t$ in $R$}: add (to $H$) a shortest path in $R$ between $p$ and $p'$. After overlaying all paths, contract away degree-2 vertices.
\end{quote}
The resulting DAM has small distortion --- one can argue that the relevant canonical pairs are enough to preserve distances between terminals: this is formalized in \Cref{lem:rel-pairs} (the ``Relevant Pairs Lemma'') and \Cref{S:rel-pairs}. Unfortunately, the DAM may be too large: it could be as large as $\tilde O_\e(|T|^2)$. Before explaining how to resolve the issue, we discuss why the DAM could be large, i.e., how a large number of splitting points could arise. As in the $+\e D$ case, we imagine processing the regions in \emph{root-to-leaf order}. When we process each region $R$, imagine drawing the paths between canonical pairs in order \EMPH{from longest to shortest}. Now consider an arbitrary path $\pi$ added between some scale-$i$ canonical pair when we process region $R$. One can show that, when path $\pi$ is added, it intersects only $\tilde O_\e(1)$ other paths \emph{that are longer than $\pi$}. (Intuitively, $\pi$ can't be close to too many portals at scale $\ge i$; this follows from the fact that separator paths are shortest paths.) One might hope that this is enough for some charging argument to bound the size of $H$. However, $\pi$ could intersect an unbounded number of paths \emph{shorter than $\pi$}; although we add paths from longest-to-shortest within region $R$, we may have added many short paths to our DAM already when an ancestor region of $R$ was processed. See \Cref{fig:bad-grid} for an example.

\paragraph{The key idea: detour paths.} Instead of adding the shortest path between a canonical pair of portals, we add an \emph{approximate} shortest path that intersects only $\tilde O_\e(1)$  paths added earlier.
Let $(p,p')$ be a canonical pair in region $R$, and initialize $\pi$ to be the shortest path between $p$ and $p'$ in $R$. As discussed above, $\pi$ can only intersect $\tilde O_\e(1)$ earlier paths which are longer than $\pi$; we just need to deal with the case where $\pi$ intersects many short paths added earlier.
Here is the key idea: Suppose that $\pi$ intersects more than $c \cdot O(\log n)$ short paths, for some number $c > 1$. Each of these short paths is a path between some canonical pair in a region above $R$, meaning that each short path has one endpoint on some external separator of $R$.  As there are only $O(\log n)$ external separators of $R$, pigeonhole principle implies that $c$ of these short paths all touch the same external separator path, which we'll call $P$. Intuitively, this means that $\pi$ stays close to $P$ for a long time. So, we modify $\pi$ by \EMPH{detouring} it along $P$: whenever $\pi$ is sufficiently close to $P$, we update $\pi$ to walk along $P$ instead of nearby. By repeatedly detouring $\pi$ against any external separators it comes too close to, we construct a path $\pi$ that can be divided into two types of subpaths: some subpaths of $\pi$ lie on the external separators of $R$, and the other subpaths of $\pi$ (which are contained in the interior of $R$) intersect only $\tilde O_\e(1)$ short paths. 
We call these the \EMPH{unsafe} and \EMPH{safe} subpaths, respectively. (Here we gloss over many subtle technical details%
; see \Cref{lem:safe} --- the ``Detour Lemma'' --- and its proof in Section~\ref{S:safe}, and an example of detouring in \Cref{fig:detour}.)

We need one last step to complete our DAM. We can add (to our DAM) the safe subpaths of $\pi$, but unfortunately the unsafe subpaths (i.e.\ the subpaths which lie on external separators) could still intersect an arbitrary number of earlier paths. 
One possible way to fix this is: for every unsafe subpath, recursively find some safe detour path. The depth of this recursion is $O(\log n)$, as each time we recurse we move upwards in the separator hierarchy. However, a naive (black-box) recursion does not work: each recursive call accumulates a multiplicative $\tilde O_\e(1)$ factor in the number of intersections, which blows up exponentially. Instead of recursing on each subpath independently, we keep track of all subpaths simultaneously during the recursion. Intuitively, if we were to detour two different subpaths of $\pi$ against the same external separator $S$, we instead only make a single detour against $S$ rather than touching $S$ twice. By doing this, we can guarantee that each level of the recursion contributes only an additive $+ \tilde O_\e(1)$ number of intersections; see \Cref{lem:impl} (the ``Proxy Lemma'') and its proof in \Cref{S:proxy}. Putting everything together, we have found an approximate shortest path $\pi$ that intersects only $\tilde O_\e(1)$ earlier paths, and so our earlier algorithm gives an $\tilde O_\e(|T|)$-size DAM.

\paragraph{Generalization and fast construction.} To conclude, we discuss two strengthenings of the strategies sketched so far. 
(1) While we have discussed the DAM for planar graphs, the approach in fact generalizes to minor-free graphs as it depends only on the existence of shortest-path separators. Note that there are several technical obstacles that arise for minor-free graphs compared to planar graphs: in planar graphs we may assume that the separator paths are shortest paths \emph{with respect to the entire graph $G$}, but in minor-free graphs these paths might only be shortest \emph{with respect to a region $R$}.
(2) In minor-free graphs, the algorithm sketched to construct DAM could require some large polynomial amount of time (at least quadratic). This is because the only construction of shortest-path separator in minor-free graphs goes through Robertson-Seymour structure theorem, and consequently state-of-the-art algorithms require at least quadratic time. Nevertheless, we are able to prove a \emph{near-linear} $\tilde O_\e(n)$ runtime on the construction of an $\tilde O_\e(|T|)$-size emulator, with a generic {bootstrapping} argument.

\paragraph{Paper Outline.}
After introducing some preliminaries about separators in Section~\ref{S:prelim},
we describe the construction of $(1+\e)$-DAM in Section~\ref{S:framework}, assuming the correctness of the 
Relevant Pairs Lemma 
\ref{lem:rel-pairs},
the Detour Lemma~\ref{lem:safe},
and the Proxy Lemma~\ref{lem:impl}.
We then prove the three key lemmas in Sections~\ref{S:rel-pairs}, ~\ref{S:safe}, and~\ref{S:proxy}, respectively.
In Section~\ref{S:fast} we describe a fast construction for the DAM.

\section{Preliminaries}
\label{S:prelim}

For any integer $n \in \mathbb{N}$, we let \EMPH{$[n]$} denote the set $\set{x \in \mathbb{N}: 0 \le x \le n}$.  For real numbers $z \in \R$ we write \EMPH{$[z]$} to denote $[\ceil{z}]$.
All logarithms are base 2.

\smallskip
In this paper we consider \emph{weighted}, \emph{undirected} graphs $G$, with vertex set \EMPH{$V(G)$} and edge set \EMPH{$E(G)$}. We use \EMPH{$\dist_G(u,v)$} to denote the distance between vertices $u$ and $v$
according to the shortest-path metric in $G$. If $u$ is a vertex and $S$ is a set of vertices, we define $\dist_G(u, S) = \min_{v \in S} \dist_G(u,v)$; for two vertex sets $S_1$ and $S_2$, we define $\dist_G(S_1, S_2) = \min_{a \in S_1, b\in S_2}(a,b)$.
If $S$ is a set of vertices, we sometimes abuse notation and write \EMPH{$\dist_S(\cdot, \cdot)$} to mean \EMPH{$\dist_{G[S]}(\cdot, \cdot)$}, that is, distance in the subgraph of $G$ induced by $S$.
We use \EMPH{$\spath_G(u,v)$} to denote the shortest path between $u$ and $v$ in the $G$. 
We use the word \EMPH{path} to denote a sequence of (possibly repeating) edges; a \EMPH{simple} path does not repeat vertices or edges. If $P$ is a path, and $a$ and $b$ are vertices on $P$, then \EMPH{$P[a:b]$} denotes the subpath of $P$ that starts at $a$ and ends at $b$. The \EMPH{diameter} of $G$ is $\max_{u,v} \dist_G(u,v)$, and we frequently denote it by \EMPH{$\spread$}. Note that if \EMPH{$\Phi$} is the maximum edge weight in $G$, and $G$ is connected, then $\spread \le (n-1)\Phi$; in particular, $\poly(\log D) = \poly(\log n, \log \Phi)$.
For every graph in this paper, we assume that the smallest edge has weight 1.

When a graph $G$ with $n$ vertices and maximum edge weight $\Phi$ is clear from context, we use the notation \EMPH{$\tilde O(x)$} to mean $x \cdot \poly(\log n, \log \Phi)$, and we use $\tilde O_\e(x)$ to mean $x \cdot \poly(\log n, \log \Phi, \e^{-1})$.

\paragraph{Unique path lengths.} 
We say that a graph $G$ has \EMPH{unique path lengths} if no two simple paths in $G$ have the same length.
In several of our constructions, we enforce this assumption by preprocessing the graph to perturb every edge weight by a small amount; specifically, we rescale all edge weights so the smallest weight is $1$, then we traverse the edges in arbitrary order and increase the weight of the $i$th edge by $+\e \cdot 2^{-i}$, where $\e$ is an arbitrary value between $0$ and~$1$. Actually, we don't have to choose a specific value of $\e$ but can instead consider the limiting behavior as $\e$ approaches 0: for every edge, store both the original weight and the value of the perturbation, and compare edges (and paths) lexicographically. (See \cite{cabello2013multiple} for details and references within.)
This transformation guarantees unique path lengths.
The perturbations require $O(n)$ bits to record, so in particular the new edge weights can be represented in $O(n)$ words\footnote{We work in the Word RAM model and assume that the edge weights of the \emph{input} graph can be stored in a single word.}, leading to only polynomial-time blowup in the runtime of our algorithms for constructing an emulator. 
The runtime blowup is the reason that we only do this preprocessing for constructions where we only care about getting polynomial time (rather than linear or near-linear time). 

We remark that, if one wanted a more time-efficient reduction, one could try to adapt the Isolation Lemma \cite{mulmuley1987matching}. If one \emph{randomly} perturbs every edge of $G$ by some random value in the range $(\e/n^4, \e)$, one can guarantee that (1) distances are distorted by a factor at most $(1+\e)$, and (2) the \emph{shortest paths} of $G$ have unique lengths \cite{cabello2013multiple}. The new edge weights can still be represented in a single word. However, this is not sufficient for our purposes; during our constructions, we will consider various different subgraphs of $G$, and we want to enforce that all shortest paths \emph{with respect to one of these subgraphs} have unique lengths. As such, any approach based on Isolatation Lemma would require more care, and so for simplicity we just use the perturbation scheme described above.


\paragraph{Shortest-path separator hierarchy.} 
A classic result of Lipton and Tarjan~\cite{lt-stpg-1979} shows that planar graphs admit shortest-path separators. 
Abraham and Gavoille~\cite{ag-olups-2006} showed that a similar result extends to minor-free graphs. Here, we introduce an alternative phrasing of their results.
A \EMPH{separator hierarchy} is a tree $\cT$ such that:
\begin{itemize}
    \item The nodes of $\cT$ are induced subgraphs \EMPH{$R$} of $G$ called \EMPH{regions}. Each region is associated with a single path \EMPH{$S$} in $R$ called an \EMPH{internal separator path}.
    The set of internal separators of nodes $R'$ that are proper ancestors of $R$ are called the \EMPH{external separators} of $R$. 
    (Note that an external separator might not contain any vertices incident to $R$.)
    \item The internal separator path $S$ of $R$ is a shortest path in $R$.%
    \footnote{Note that, unlike in planar graphs, the separator may not be a shortest path in $G$; it is only a shortest path in $R$.}
    \item The regions of the children of $R$ are the connected components of $R \setminus S$. In particular, if $R \setminus S = \varnothing$, then $R$ is a leaf of $\cT$.
    \item Any path in $G$ that contains a vertex in $R$ and a vertex not in $R$ must also contain a vertex in an external separator of $R$.
\end{itemize}
The \EMPH{height} of the separator hierarchy is the height of the tree $\cT$.
We say that a region $R$ is \EMPH{above} (resp.\ \EMPH{below}) a region $R'$ if $R$ is an ancestor (resp.\ descendant) of $R'$ in the separator hierarchy. Every node is an ancestor and descendant of itself, so each region $R$ is both above and below itself.

\begin{lemma}[\cite{ag-olups-2006}]
    Every $K_r$-minor-free graph admits a separator hierarchy of height $O_r(\log n)$.%
    \footnote{Those who are familiar with the work by Abraham and Gavoille~\cite{ag-olups-2006} may notice the slight difference in our definition.  In particular, they iteratively remove separators which are the \emph{union} of $k = O_r(1)$ shortest paths.
    We can recover our version of separator hierarchy from theirs by 
    splitting up the $k$ shortest paths in their separator into $O(k^2)$ \emph{disjoint} shortest paths (because any pair of shortest paths can only meet and split apart one time, by our assumption that $G$ has unique path lengths), and replace the node in $\cT$ with a sequence of nodes that removes these paths one at a time.
    }
\end{lemma}





\paragraph{Portals.}
We define \emph{portals} along each internal separator path, analogous to the $\e$-cover of \cite{Thorup04}.
We fix some $\e \in (0,1)$, specified later.
Let $R$ be a region, and let $S$ be the internal separator path of $R$. For every $i \ge 1$, we now define \EMPH{scale-$i$ portals in $R$} --- for short, the \EMPH{$(i, R)$-portals}. For $i = 1$, we define the $(1, R)$-portals to be the set of all vertices on $S$. For $i > 1$, we define the $(i, R)$-portals to be a set \EMPH{$\Pi_i$} which is a subset of the $(i-1, R)$-portals $\Pi_{i-1}$, satisfying:
\begin{enumerate}
    \item[(1)] every vertex $s$ in $\Pi_{i-1}$ satisfies $\dist_{S}(s, \Pi_i) \le \frac{\e}{2} \cdot 2^i$, and
    \item[(2)] every pair of vertices $p_1, p_2 \in \Pi_i$ satisfies $\dist_S(p_1, p_2) \ge \frac{\e}{2} \cdot 2^i$.
\end{enumerate}
Such a portal set can be constructed greedily.
The \EMPH{scale-$i$ portals} refer to the set of all $(i,R)$-portals over all regions $R$ in $\cT$.
We remark that portals at scale $i$ will be responsible for preserving distances around $2^i$, up to a $(1+\e)$ factor. As such, they are separated by distance roughly $\e \cdot 2^i$ (not by distance $2^i$).
A geometric series arising from property (1) of $(i, R)$-portals leads to the following property:
\begin{observation}
    Let $R$ be a region with internal separator $S$, and let $i$ be a scale. Let $\Pi_i$ denote the scale-$i$ portals of $R$. Every vertex $s \in S$ satisfies $\dist_S(s, \Pi_i) \le \e 2^i$.
\end{observation}

The following claim bounds the number of portals that are close to any fixed vertex.
\begin{restatable}{claim}{lengththreat}
\label{clm:length-threat}
    Let $R$ be a region and let $i \in \mathbb{N}$ be a scale.
    Let $\alpha = O(1)$ be any constant. 
    \begin{itemize}
        \item For any vertex $v$ in $R$, the number of $(i, R)$-portals $p$ with $\dist_{R}(v, p) \le \alpha \cdot 2^i$ is $O(\alpha \cdot \e^{-1})$.
        \item For any path $\pi$ in $R$, the number of $(i, R)$-portals $p$ with $\dist_{R}(\pi, p) \le \alpha \cdot 2^i$ is $O(\alpha \cdot \e^{-1} \cdot \ceil{\frac{\norm{\pi}}{2^i}})$.
    \end{itemize}
\end{restatable}

\begin{proof}
We prove the second statement, which implies the first statement (because any vertex $v$ is a path with length 0).
Let $S$ be the internal separator of $R$. Let $\Pi = \set{p_1, \ldots, p_\beta}$ be the set of $(i, R)$-portals $p$ such that $\dist_{R}(p, \pi) \le \alpha \cdot 2^i$. Notice that every $(i, R)$-portal lies on the path $S$.
Without loss of generality assume that $p_1$ is the first portal in $\Pi$ when walking along $S$ (in an arbitrary direction), and that $p_\beta$ is the last portal in $P$ when walking along $S$.
By property (2) in the definition of portals, we have $\dist_{S}(p_1, p_\beta) \ge \frac{\e}{2} \cdot 2^i \cdot (\beta - 1)$; because $S$ is a shortest path in $R$, we further have 
\[
\dist_{R}(p_1, p_\beta) \ge \frac{\e}{2} \cdot 2^i \cdot (\beta - 1).
\]
Because $\pi$ is a path in $R$, triangle inequality implies that
\[\dist_{R}(p_1, p_\beta) \le \dist_{R}(p_1, \pi) + \norm{\pi} + \dist_{R}(\pi, p_\beta) \le 2 \alpha \cdot 2^i + \norm{\pi}. \]
Combining these two inequalities, we find that $\beta = O(\alpha \cdot \e^{-1} \cdot \ceil{\frac{\norm{\pi}}{2^i}} )$ as claimed.
\end{proof}

\section{A framework for $(1+\e)$-approximate distance-approximating minor}
\label{S:framework}

In this section, we construct a $(1+\e_0)$-approximate distance-approximating minor for any planar graph, with some $\log$ dependence on the size of the graph and the maximum edge weight.

\begin{restatable}{theorem}{damScaled}
\label{thm:dam-scaled}
Let $G$ be an $n$-vertex minor-free graph, with edge weights between $1$ and $\Phi$. Let $T$ be a set of terminals in $G$, and let $\e_0$ be a parameter in $(0,1)$.
There is a $(1+\e_0)$-approximate DAM for $G$ with respect to the terminals $T$, with size $O(\e_0^{-7} \cdot (\log n)^{30} \cdot (\log n\Phi)^{13} \cdot |T|)$. The DAM can be constructed in $\poly(n, \log \Phi, \e^{-1})$ time.
\end{restatable}
We assume WLOG that $G$ is connected (otherwise we construct a DAM for each connected component individually). Let \EMPH{$\spread$} denote the diameter of $G$, and observe that $\spread \le n\Phi$. We aim to construct a DAM with size $O(\e_0^{-7} \cdot (\log n)^{30} \cdot (\log \spread)^{13} \cdot |T|)$.
\paragraph{Rescaling \boldmath{$\e$}.}
A subtle details about setting the error parameter $\e$ is that in the definition of scale-$i$ portals (see \Cref{S:prelim}), we implicitly assume $\e$ to be a fixed constant.  
This means that we won't be able to rescale $\e$ from this point on. 
Instead, we will work with a ``scaled down'' version of $\e$ in the next two sections of this paper for the purpose of proving Theorem~\ref{thm:dam}.
We set $\EMPH{$\e$} \coloneqq \frac{\e_0}{\Theta(\log^3 n \cdot \log \spread)}$ with some sufficiently large constant in the denominator, and define portals in terms of this $\e$. Showing a $1 + O(\log^3 n \cdot \log \spread) \cdot \e$ stretch will then provide the final $(1+\e_0)$-approximation as desired.
%
%
%


\paragraph{Runtime.} It will be clear from our proofs that our DAM can be constructed in time $\poly(n, \log \Phi, \e^{-1})$. To streamline the presentation, we do not dwell on the proof of the runtime bound and instead focus on the existence of a DAM. All of our algorithms proceeds in $O(\log \Phi)$ iterations, where each iteration computes and manipulates some shortest paths in $\poly(n, \e^{-1})$ time.

\subsection{Ingredients for constructing a DAM: Canonical pairs and safe paths}
\label{SS:ingredients}

Our goal for this subsection is to define a set of $\tilde{O}(n)$ ``safe paths'' that intersect sparsely with other safe paths; in the next subsection, we will construct our emulator by taking the union of some subset of safe paths.  
A \EMPH{canonical pair at scale $i$ in region $R$}, also called an \EMPH{$(i, R)$-canonical pair} for short, is a pair of vertices denoted \EMPH{$\canon a b$}
where (1) $a$ is a scale-$i$ portal on the internal separator of $R$, and $b$ is a scale-$i$ portal on some external or internal separator $S$ of $R$; and (2) $\dist_{R^{\arcto}}(a,b) \le 2^i$, where $\EMPH{$R^{\arcto}$}$ is the induced subgraph of $G$ which is induced by the vertices in $R \cup S$\footnote{We remark that if $S$ is an internal separator of $R$, then $S \subseteq R$ and so $R^{\arcto}$ is the same as $R$. But if $S$ is an external separator of $R$, then $S$ is disjoint from $R$, and $R^{\arcto}$ is different than $R$.}.
The subgraph $R^{\arcto}$ is called the \EMPH{canonical subgraph} for $\canon{a}{b}$.
We emphasize that not every scale-$i$ portal on $S$ is  directly adjacent to $R$, but nevertheless we can define $\canon a b$.
    
A \EMPH{$t$-canonical sequence} between vertices $a$ and $b$, with respect to subgraph $H$, is a sequence of vertices $(x_1, \ldots, x_k)$ such that
    (1) $x_1 = a$ and $x_k = b$; 
    (2) for every $i \in \set{1,\ldots,k-1}$, either $\canon {x_i} {x_{i+1}}$ or $\canon {x_{i+1}} {x_i}$ is a canonical pair with some canonical subgraph $R_i^{\arcto}$; and
    (3) $\sum_{i=1}^{k-1} \dist_{R_i^{\arcto}}(x_i, x_{i+1}) \le t \cdot \dist_H(a,b)$.
We can think of a canonical sequence between $a$ and $b$ as defining a $t$-approximate shortest path between $a$ and $b$ in $H$, where this approximate path is composed of shortest paths between vertices of canonical pairs.
We will mainly work with $t$-canonical sequences \EMPH{with respect to $G$}. We often omit the words ``with respect to $G$''; i.e., a canonical sequence defaults to being with respect to $G$, and we only specify $H$ when $H \neq G$.

\begin{lemma}[Relevant Pairs Lemma]
\label{lem:rel-pairs}
Every vertex $v$ in $G$ can be associated with a set of canonical pairs, called the \EMPH{relevant canonical pairs} and denoted \EMPH{$\relP(v)$}, such that for every pair of vertices $(a,b)$,
there is a $(1 + O(\log n)\cdot\e)$-canonical sequence
between $a$ and $b$ such that every canonical pair in the sequence is in $\relP(a) \cup \relP(b)$.
Further, there are at most $O(\e^{-2} \cdot \log^2 n \cdot \log^2 \spread)$ pairs in $\relP(v)$ per vertex $v$.
\end{lemma}

We prove Lemma~\ref{lem:rel-pairs} in Section~\ref{S:rel-pairs}.
The lemma shows that preserving distances just between relevant canonical pairs suffices to preserve all pairwise terminal-to-terminal distances. (We remark that, although our precise statement of \Cref{lem:rel-pairs} is new, similar ideas have appeared in previous literature, e.g. \cite{acg-fdado-2012}.) As described in the technical overview, Lemma~\ref{lem:rel-pairs} leads to a $(1+\e)$-approximate \emph{non-planar emulator} of near-linear size.
To obtain our DAM, we need to find $(1+\e)$-approximate paths between canonical pairs, which (in some amortized sense) have few intersections with the approximate paths added for other canonical pairs. To make this precise, we introduce two definitions.

%

\begin{definition}
\label{def:threatening}
    Let $\pi$ be a path. 
    Let $\canon{a}{b}$ be an $(i, R)$-canonical pair. We say that $\pi$ \EMPH{threatens} $\canon a b$ if $\dist_{R}(\pi, a) \le 2 \cdot 2^i$.%
\end{definition}

To aid intuition, note that if $\pi$ threatens $\canon a b$, then $\pi$ could potentially intersect a $(1+\e)$-approximate shortest path in $R$ between $a$ and $b$.
A priori, a shortest path $\pi$ could threaten many canonical pairs.

\begin{definition}
    Define a \EMPH{partial order $\preceq$} on canonical pairs as follows. If $\canon a b$ is an $(i, R)$-canonical pair, and $\canon {a'} {b'}$ is an $(i', R')$-canonical pair, we say $(\canon a b) \preceq (\canon {a'}{b'})$ if either (1) $R'$ is a proper ancestor of $R$ in the separator hierarchy, or (2) $R = R'$ and $i \le i'$.  
    In short, the partial order $\preceq$ is a lexicographical order of canonical pairs based on region containment followed by scale.
\end{definition}


\medskip
\noindent The following is a key lemma, which we prove in the Section~\ref{S:safe}.

\begin{lemma}[Detour Lemma]
\label{lem:safe}
    %
    For every canonical pair $\canon a b$ in region $R$ with canonical subgraph $R^{\arcto}$, there exists a \EMPH{detour path} $P$ in $G$
    such that:
    \begin{itemize}
        \item \textnormal{[Stretch.]} 
        The length of $P$ is at most $(1+ O(\log n \cdot \log \spread) \cdot \e) \cdot \dist_{R^{\arcto}}(a,b)$.
        \item \textnormal{[Subpath decomposition.]} 
        The edges of $P$ can be partitioned into a collection $\cP$ of $O(\log n \cdot \log \spread)$ subpaths, such that each subpath in $\cP$ is either (1) a single edge (between $R$ and an external separator of $R$,
        (2) a shortest path in $R$, or (3) a subpath of an external separator of $R$. These subpaths are called, respectively, the \EMPH{safe edges}, the \EMPH{safe subpaths}, and the \EMPH{unsafe subpaths}.
        Let \EMPH{$\safe(\canon a b)$} denote a set of safe edges and safe subpaths of the detour path for canonical pair $\canon a b$.
    \end{itemize}
    Every safe subpath $\pi$ satisfies the following properties:
    \begin{itemize}
        \item \textnormal{[Bounded threats.]} 
        $\pi$ threatens only $O(\e^{-4}\log^3 n \log \spread)$ canonical pairs $\canon {a'} {b'}$ with $(\canon a b) \preceq (\canon {a'} {b'})$.
        \item \textnormal{[Bounded splits.]} 
        Let $\pi'$ be a shortest path in some region $R'$ that is above $R$.
        We say a vertex $v$ is a \EMPH{splitting point} of $\pi$ and $\pi'$ if it has degree larger than 2 in the subgraph $\pi \cup \pi'$. There are at most $O(\e^{-1}\log n)$ splitting points of $\pi$ and $\pi'$.
    \end{itemize}
\end{lemma}

When constructing our emulator, if we want to preserve the distance between $a$ and $b$, we can afford to add the subpaths of $\safe(\canon a b)$ to the emulator (because the safe subpaths threaten only $\tilde{O}_\e(1)$ portals, and the safe edges can be dealt with separately), but we cannot afford to add the unsafe subpaths. As such, we need one more lemma before we can construct our emulator, which we prove in Section~\ref{S:proxy}.

\begin{lemma}[Proxy Lemma]
\label{lem:impl}
    For every canonical pair $\canon a b$ with canonical subgraph $R^{\arcto}$, there is a set \EMPH{$\proxy(a,b)$} containing $\canon a b$ 
    and other $O(\log n \cdot \log \spread)$ canonical pairs such that the subgraph $H$ that is the union of all the paths in
    \[
    \bigcup_{\canon {a'} {b'} \in \proxy(a,b)} \safe(\canon {a'} {b'})
    \]
    satisfies $\dist_{H}(a,b) \le (1+ O(\log^2 n \cdot \log \spread) \cdot \e) \cdot \dist_{R^{\arcto}}(a,b)$.
\end{lemma}
The intuition is that the safe subpaths of $\proxy(a,b)$ serve as a replacement for the unsafe paths of $(a,b)$: even though the graph $H$ does not include the unsafe subpaths for $\canon a b$, the safe subpaths of the proxy pairs can be used to bypass these unsafe subpaths.

\subsection{Constructing a distance-approximating minor}
We now prove \Cref{thm:dam-scaled} (restated here), assuming the lemmas stated in Section~\ref{SS:ingredients}. Recall that we set $\e \coloneqq \frac{\e_0}{\Theta(\log^3 n \cdot \log \spread)}$, and that $D \le n\Phi$.

\damScaled*

\begin{tcolorbox}[boxsep=0mm,left=4mm,right=4mm,leftrule=2mm]
%
\paragraph{DAM Construction.}
Perturb the edge weights so that $G$ has unique path lengths.
For each terminal $t$ in $T$, define a set \EMPH{$\cP_t$} of paths as follows:
for every canonical pair $\canon a b$ in $\relP(t)$, for every pair $\canon {a'} {b'}$ in $\proxy(a,b)$, for every safe subpath and safe edge $\pi$ in $\safe(\canon {a'} {b'})$, add $\pi$ to the set $\cP_t$.  
Let \EMPH{$\cP$} denote the set $\bigcup_{t \in T} \cP_t$.
Let \EMPH{$M$} be the subgraph obtained by taking the union of all paths in $\cP$. Let \EMPH{$\check M$} be the subgraph $M$ after contracting away all non-terminal degree-2 vertices; an edge between two vertices $u$ and $v$ in $\check M$ is given the weight $\dist_{M}(u,v)$.
\end{tcolorbox}

\smallskip
\noindent We claim that $\check M$ is a $(1+\e_0)$-DAM with size $O(|T| \cdot \poly(\log n, \log \spread, \e_0^{-1}))$. 

\begin{lemma}
\label{lem:dam}
    $\check M$ is a $(1+\e_0)$-distance-approximating minor for $T$.
\end{lemma}

\begin{proof}
    Clearly $\check M$ is a minor of $G$, as it is obtained by contracting edges of a subgraph of $G$. Now, let $t_1$ and $t_2$ denote two terminals in $T$. Our choice of edge weights of $\check M$ guarantees that distances are non-contracting, i.e. $\dist_G(t_1, t_2) \le \dist_{\check M}(t_1, t_2)$.
    We now show that $\dist_{\check M}(t_1, t_2) \le (1+O(\log^3 n \cdot \log \spread)\cdot \e) \cdot \dist_G(t_1, t_2)$; the lemma follows by definition of $\e$.
    
    By Lemma~\ref{lem:rel-pairs}, there exists a canonical sequence of vertices $(v_0, \ldots, v_k)$ such that 
    (1) $v_0 = t_1$ and $v_k = t_2$; 
    (2) for every $i \in \set{0, \ldots, k-1}$, either $\canon{v_i}{v_{i+1}}$ or $\canon{v_{i+1}}{v_{i}}$ is in $\relP(t_1) \cup \relP(t_2)$ with some canonical subgraph $R^{\arcto}_i$; and
    (3) $\sum_{i = 0}^{k-1} \dist_{R^{\arcto}_i}(v_i, v_{i+1}) \le (1+O(\log n)\cdot \e) \cdot \dist_G(t_1, t_2)$.
    It follows from the construction of $M$ and condition (2) that, for every $i$, subgraph $M$ either contains every path in 
    \[
    \bigcup_{\canon {a'} {b'} \in \proxy(v_i,v_{i+1})} \safe(\canon {a'} {b'})
    \]
    or every path in
    \[
    \bigcup_{\canon {a'} {b'} \in \proxy(v_{i+1},v_{i})} \safe(\canon {a'} {b'}),
    \]
    depending on whether $\canon {v_i}{v_{i+1}}$ or $\canon{v_{i+1}} {v_i}$ is in $\relP(t_1) \cup \relP(t_2)$.
    Either way, Lemma~\ref{lem:impl} implies that $\dist_M(v_i, v_{i+1}) = \dist_M(v_{i+1}, v_i) \le (1+ O(\log^2 n \cdot \log \spread)\cdot \e) \cdot \dist_{R^{\arcto}_i}(v_i, v_{i+1})$. 
    We conclude 
    \begin{align*}
    \dist_M(t_1, t_2) &\le \sum_{i = 0}^{k-1} \dist_M(v_i, v_{i+1}) \\
    &\le (1+O(\log^2 n \cdot \log \spread)\cdot \e) \cdot \sum_{i=0}^{k-1}  \dist_{R^{\arcto}_i}(v_i, v_{i+1}) \\
    &\le (1+O(\log^3 n \cdot \log \spread)\cdot \e) \cdot \dist_G(t_1, t_2).
    \end{align*}
    Distances between terminals in $\check M$ are the same as in $M$, so $\dist_{\check M}(t_1, t_2) \le (1+O(\log^3 n \cdot \log \spread))~\cdot~\dist_G(t_1,t_2)$.
\end{proof}

We now bound the size of $\check M$. 
Recall that if $\pi_1$ and $\pi_2$ are paths, we say that a vertex $v$ is a \emph{splitting point} of $\pi_1$ and $\pi_2$ if $v$ has degree greater than 2 in the graph $\pi_1 \cup \pi_2$. Observe that the size of $\check M$ is bounded by the number of splitting points between all pairs of paths in $\cP$ (see Lemma~\ref{lem:size}). We first prove a lemma that lets us ``charge'' every splitting point to a threatening subpath (Lemma~\ref{lem:intersect-implies-threat}), and use this to prove the size bound (Lemma~\ref{lem:size}).

\begin{lemma}
\label{lem:intersect-implies-threat}
    Let $\pi_j$ be a subpath in $\safe(\canon {a_j} {b_j})$, where $\canon {a_j} {b_j}$ is some $(i_j, R_j)$-canonical pair with canonical subgraph $R^{\arcto}_j$, for both $j \in \set{1,2}$.  
    Suppose $\pi_1$ and $\pi_2$ intersect at some vertex that is not an endpoint of $\pi_1$ or $\pi_2$. Then without loss of generality $(a_1, b_1) \preceq (a_2, b_2)$ 
    and $\pi_1$ threatens $\canon{a_2}{b_2}$.
\end{lemma}

\begin{proof}
Let $v$ be some vertex in both $\pi_1$ and $\pi_2$, such that $v$ is not an endpoint of $\pi_1$ or $\pi_2$.
Because $v$ exists, paths $\pi_1$ and $\pi_2$ are both longer than a single edge; this, together with the definition of $\mathrm{Safe}(\cdot)$ (from Lemma~\ref{lem:safe}) implies that $\pi_1$ (resp.\ $\pi_2$) is contained in $R_1$ (resp.\ $R_2$) and thus $v$ is in $R_1 \cap R_2$.
By construction of the separator hierarchy, $R_1 \cap  R_2$ is non-empty only if $R_1$ and $R_2$ are in an ancestor-descendant relation.
We assume without loss of generality that $R_2$ is an ancestor of $R_1$; if $R_1 = R_2$, then we assume without loss of generality that $i_1 \le i_2$. In other words, we assume without loss of generality that $(a_1, b_1) \preceq (a_2, b_2)$.

To show that $\pi_1$ threatens $\canon{a_2}{b_2}$, we must show that $\dist_{R^{\arcto}_2}(\pi_1, a_2) \le 2\cdot 2^{i_2}$.
Observe that $\pi_2$ is a path in $R_2 \subseteq R^{\arcto}_2$. Moreover, $\pi_2$ has length at most $(1+O(\log n \cdot \log \spread)\cdot \e) \cdot 2^{i_2} \le 2 \cdot 2^{i_2}$; this follows from the [stretch] property of Lemma~\ref{lem:safe} (as $\pi_2$ is a subpath of the detour path of a scale-$i_2$ canonical pair), and from the fact that $\e = \frac{\e_0}{\Theta(\log^3 n \cdot \log \spread)}$ (and so $O(\log n \cdot \log \spread)\cdot \e \le 1$). Thus, $\dist_{R^{\arcto}_2}(v, a_2) \le 2 \cdot 2^{i_2}$ and, because $v$ is also in $\pi_1$, we conclude $\dist_{R^{\arcto}_2}(\pi_1, a_2) \le 2 \cdot 2^{i_2}$.
\end{proof}

\begin{lemma}
\label{lem:size}
    $\check M$ has $O(\e_0^{-7} \log^{30} n \log^{13} \spread) \cdot |T|)$ vertices and edges. 
\end{lemma}

\begin{proof}
    For every terminal $t$, there are $O(\e^{-2} \cdot \log^2 n \cdot \log^2 \spread)$ canonical pairs in $\relP(t)$ (by Lemma~\ref{lem:rel-pairs}); 
    for each canonical pair $\canon a b$ in $\relP(t)$, there are ${O}(\log n \cdot \log \spread)$ canonical pairs $\canon {a'}{b'}$ in $\proxy(a,b)$ (by Lemma~\ref{lem:impl}); 
    and for each such canonical pair $\canon{a'}{b'}$, there are ${O}(\log n \cdot \log \spread)$ subpaths in $\safe(\canon {a'} {b'})$ (by Lemma~\ref{lem:safe} [subpath decomposition]).
    In total, $M$ is the union of ${O}( \e^{-2}\log^4 n \log^4\spread \cdot |T|)$ paths.

    As $\check M$ is minor-free, it suffices to bound the number of vertices. The number of vertices in $\check M$ is precisely the number of vertices in $M$ with degree not equal to 2. 
    If a vertex in $M$ has degree not equal to 2, either it is the endpoint of a path in $\cP$, or it is a splitting point of two paths in $\cP$. 
    There are $O(|\cP|) = O(\e^{-2}\log^4 n \log^4\spread \cdot |T|)$ vertices of the first type (endpoints).  
    It remains to bound the number of vertices of the second type (splitting points).
    
    We use a charging argument.
    Let $v$ be a splitting point of two paths $\pi$ and $\pi'$ in $\cP$, where $v$ is not an endpoint of $\pi$ or $\pi'$. Both $\pi$ and $\pi'$ are safe subpaths (and not safe edges) of some canonical pair. Let $\canon a b$ be the canonical pair such that $\pi \in \safe(\canon a b)$, and $\canon {a'} {b'}$ be the canonical pair where $\pi' \in \safe(\canon {a'} {b'})$; 
    without loss of generality, by Lemma~\ref{lem:intersect-implies-threat}, we have $(\canon a b) \preceq (\canon {a'}{b'})$ and $\pi$ threatens $\canon {a'} {b'}$. We charge the splitting point $v$ to the threatening path $\pi$.
    An arbitrary safe subpath $\pi \in \cP$ (where we let $\canon a b$ denote the canonical pair with $P \in \safe(\canon {a} {b})$) 
    receives at most one charge for every splitting vertex between $\pi$ and a path $\pi'$, for every safe subpath $\pi'$ in $\safe(\canon {a'} {b'})$, for every canonical pair $\canon {a'} {b'}$ with $(\canon {a} {b}) \preceq (\canon {a'} {b'})$ threatened by $\pi$.
    \begin{itemize}
        \item The safe subpath $\pi$ threatens at most ${O}(\e^{-4}\log^3 n \log \spread)$ canonical pairs $\canon {a'} {b'}$ with $(\canon {a} {b}) \preceq (\canon {a'} {b'})$, by Lemma~\ref{lem:safe} [bounded threats].
        \item For each fixed $\canon {a'} {b'}$,
        there are ${O}(\log n \log \spread)$ paths $\pi' \in \safe(\canon {a'} {b'})$, again by Lemma~\ref{lem:safe} [subpath decomposition].
        \item For each fixed $\canon {a'} {b'}$ and $\pi' \in \safe(\canon {a'} {b'})$,
        there are $O(\e^{-1} \log n)$ splitting points between $\pi$ and $\pi'$. Indeed, $\pi'$ is a shortest path in the region $R'$ associated with $\canon {a'} {b'}$, by Lemma~\ref{lem:safe} [subpath decomposition]. As $(\canon {a} {b}) \preceq (\canon {a'} {b'})$, region $R'$ is a (not necessarily proper) ancestor of the region $R$ associated with $\canon a b$. Thus, Lemma~\ref{lem:safe} [bounded splits] implies that there are ${O}(\e^{-1} \log n)$ splitting points between $\pi$ and $\pi'$. 
    \end{itemize}
    Thus, the path $\pi$ has charge at most ${O}(\e^{-5} \log^5 n \log^2 \spread)$. 
    The total charge over all paths $\pi \in \cP$ is ${O}(|\cP|) \cdot {O}(\e^{-5} \log^5 n \log^2 \spread) = {O}(\e^{-7} \log^{9} n \log^{6}\spread \cdot |T|)$;
    it follows that the number of non-degree-2 vertices in $M$, and thus the number of vertices in $M$, is at most ${O}(\e^{-7} \log^{9} n \log^{6}\spread \cdot |T|)$. (As $M$ is minor-free, the number of edges is linear in the number of vertices.)
    As $\e = \smash{\frac{\e_0}{O(\log^3 n \cdot \log \spread)}}$, the size of $M$ is $O(\e_0^{-7} \log^{30} n \log^{13} \spread \cdot |T|)$ as claimed.
\end{proof}

Lemmas~\ref{lem:dam}~and~\ref{lem:size} (together with the fact $D \le n \Phi$) imply Theorem~\ref{thm:dam-scaled}.


\section{Proof of Lemma~\ref{lem:safe}:  Detour paths}
\label{S:safe}
\begin{figure}
    \centering
    \includegraphics[width=0.75\linewidth]{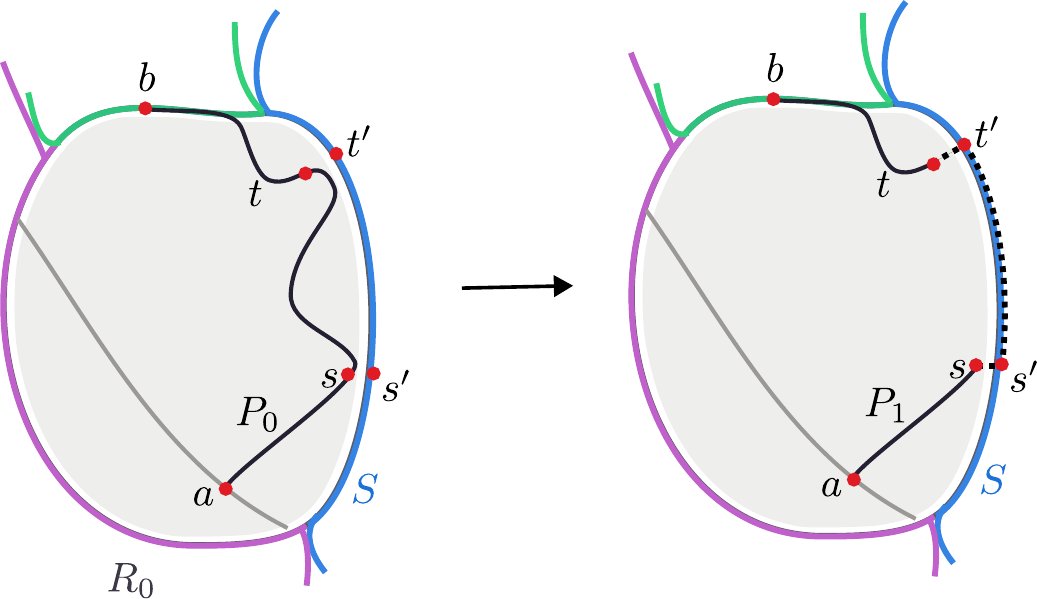}
    \caption{A stylized depiction of one iteration of \textsc{DetourPath}, during which path $P_0$ is detoured against separator $S$. On the left: There is a canonical pair $\canon a b$ in the region $R_0$; the region $R_0$ has an internal separator path drawn in gray and external separator paths drawn in colors. A shortest path $P_0$ between $a$ and $b$ in drawn in black, and vertices $s$, $s'$, $t$ and $t'$ are marked. On the right: $P_0$ is detoured against $S$ to create path $P_1$.}
    \label{fig:detour}
\end{figure}

We begin with the proof of \Cref{lem:safe}, which we view as a main technical innovation in this paper.
Let $\canon a b$ be a canonical pair with scale \EMPH{$i_0$} and region \EMPH{$R_0$}. Recall that $R_0^{\arcto}$ denotes the canonical subgraph, which is the union of $R_0$ and the separator containing $b$.
The procedure \EMPH{$\textsc{DetourPath}(\canon a b)$} described below returns a path \EMPH{$P$} satisfying the constraints of Lemma~\ref{lem:safe}. Roughly speaking, we initialize $P$ to be the shortest path from $a$ to $b$ in $R_0^{\arcto}$. Next, we consider each external separator path $S$ of $R_0$: if $P$ is within $\e \norm{P}$ of $S$, 
then we change $P$ to ``detour'' along $S$. 
That is, we let $s'$ (resp.\ $t'$) be the first (resp.\ last) vertex along $S$ where $P$ is within distance $\e \norm{P}$ of $s'$ (resp.\ $t'$), and we replace $P$ with concatenated shortest paths from $a$ to $s'$, from $s'$ to $t'$, and from $t'$ to $b$. 
After making these ``$\e$-detours'' against all the external separators of $P$, we repeat the process, but this time we only detour if $\pi$ is within distance $\frac{\e}{2} \norm{P}$ of $S$. Repeat for $\log_2 (\e\norm{P})$ rounds, shrinking the threshold for detours by a factor 2 after each round.
See Figure~\ref{fig:detour-code} for the pseudocode, and \Cref{fig:detour} for an example of one iteration.

\begin{figure}[h!]
\small
\centering
\begin{algorithm}
\internallinenumbers
\textul{$\textsc{DetourPath}(\canon a b)$}: \+
\\  $P_0 \gets $ shortest path from $a$ to $b$ in $R_0^{\arcto}$ (the canonical subgraph of $\canon a b$)
\\  $k \gets 0$ \Comment{the iteration counter}
\\  for each scale $i$ from $\lceil \log_2 (\e \norm{{P_0}}) \rceil$ down to $0$: \+
\\  for each external separator path $S$ of $R_0$:\+
\\      increase $k$ by $1$
\\~
\\      \Comment{Change $P$ by detouring along $S$ if possible}
\\      $s \gets$ first vertex along $P_{k-1}$ such that $s \in R_0$ and $\dist_{R_0 \cup S}(s, S) \le 2^i$
\\      $s' \gets$ a vertex in $S$ that minimizes $\dist_{R_0 \cup S}(s, s')$.
\\      $t \gets$ last vertex along $P_{k-1}$ such that $t \in R_0$ and $\dist_{R_0 \cup S}(t, S) \le 2^i$
\\      $t' \gets$ a vertex in $S$ that minimizes $\dist_{R_0 \cup S}(t, t')$.
\\
\\  if $s$ or $t$ do not exist:\+
\\      $P_{k} \gets P_{k-1}$\-
\\ otherwise:\+
\\      ${P}_{k} \gets {P}_{k-1}[a:s] \circ \spath_{R_0 \cup S}(s,  s') \circ \spath_{R_0 \cup S}(s', t') \circ \spath_{R_0 \cup S}(t', t) \circ {P}_{k-1}[t:b]$\-
\\  \-\-
\\  return $P_k$ as $P$
\end{algorithm}
\caption{The procedure \textsc{DetourPath}.
}
\label{fig:detour-code}
\end{figure}

Let \EMPH{$P$} denote the output of $\textsc{DetourPath}(\canon a b)$. The remainder of this section is devoted to proving the four properties required by Lemma~\ref{lem:safe}: the [subpath decomposition], [stretch], [bounded threats], and [bounded splits] properties. 

\subsection{[Subpath decomposition] property.}
The [subpath decomposition] property is relatively straightforward. By construction, $P$ is a path (though not necessarily a simple path).

\begin{lemma}[Subpath decomposition]
\label{lem:detour-subpath}
    The output path $P$ of $\textsc{DetourPath}(\canon a b)$ for any canonical pair in region $R_0$ can be partitioned into $O(\log n \cdot \log \spread)$ edge-disjoint subpaths such that each subpath is either (1) a single edge (``safe edge''), (2) a shortest path in $R_0$ (``safe subpath''), or (3) a subpath of some external separator of $R_0$ (``unsafe subpath'').
\end{lemma}

\begin{proof}
We say that a (sub)path is \EMPH{valid} if it is either a single edge, a shortest path in $R_0$, or a subpath of some external separator of $R_0$. For each path \EMPH{$P_k$} constructed during the execution of $\textsc{DetourPaths}(\canon a b)$, we (inductively) define a set \EMPH{$\cP_k$} of subpaths that partition \EMPH{$P_k$}, where each subpath in $\cP_k$ is valid and $|\cP_k| \le 7k + 3 = O(k)$. The lemma will follow from the fact that the number of iterations is $O(\log \spread \cdot \log n)$

As the base case, we construct $\cP_0$. Let \EMPH{$S_b$} denote the (internal or external) separator of $R_0$ that contains $b$. Recall that $P_0$ is a shortest path from $a$ to $b$ in $R_0^{\arcto} = R_0 \cup S_b$. Let \EMPH{$x$} be the first vertex along $P_0$ (where we view $P_0$ as starting at $a$ and ending at $b$) that is in $S_b$, and let \EMPH{$\operatorname{pred}(x)$} be the predecessor of $x$ in $P_0$. We define \EMPH{$\cP_0$} to be the set containing the subpath $P_0[a:\operatorname{pred}(x)]$, the edge $(\operatorname{pred}(x), x)$, and the subpath $P_0[x:b]$. Clearly, each subpath in $\cP_k$ is valid ---  $P_0[a:\operatorname{pred}(x)]$ is contained in $R_0$ (by the minimality of $x$), and it is a shortest path in $R_0$ because it is a subpath of a shortest path in $R_0 \cup S_b$; $(\operatorname{pred}(x), x)$ is an edge; and $P_0[x:b]$ is contained in the external separator $S_b$ (as $S_b$ is a shortest path in $R_0 \cup S_b$ that contains both $x$ and $b$, and shortest paths in $R \cup S_b$ are unique). Further, $\cP_0$ is a partition of the edges of $P_0$, and $\cP_0$ contains only $3$ subpaths.

In the inductive case, we construct $\cP_{k}$, assuming $\cP_{k-1}$ has been constructed. If $P_k = P_{k-1}$, the claim follows trivially by induction. Otherwise, $P_k$ was constructed by defining vertices $s$, $s'$, $t'$, $t$ and detouring through these vertices.
In this case, the subpaths in $\cP_{k}$ are as follows.
\begin{enumerate}
    \item If a subpath of $\cP_{k-1}$ was entirely contained in $P_k[a:s]$ or $P_k[t:b]$, it gets added to $\cP_{k}$.
    \item There is a (possibly empty) suffix of $P_k[a:s]$ that was not assigned in item (1) above (because it belonged to a subpath that was split by $s$). Assign this suffix to be a new subpath in $\cP_{k}$.  Similarly, there is a prefix of $P_k[t:b]$ that was not assigned in the first bullet point; assign this to be a new subpath in $\cP_{k}$.
    \item Let $\mathrm{pred}(s')$ be the last vertex of $\spath(s,s')$ before $s'$, and let $\mathrm{next}(t')$ be the first vertex of $\spath(t', t)$ after $t'$. 
    Add the edge $(\mathrm{pred}(s'), s')$ and the edge $(t, \mathrm{next}(t'))$ as subpaths of $\cP_{k}$
    \item Add the subpaths $\spath(s, \mathrm{pred}(s'))$ and $\spath(\mathrm{next}(t'), t')$.
    \item Add the subpath $\spath(s', t')$.
\end{enumerate}

It is straightforward to see that each subpath in $\cP_{k}$ is valid --- the subpaths added in steps 1 and 2 are (subpaths of) subpaths in $\cP_{k-1}$ and are valid by induction; the subpaths in step 3 are a single edge; the subpaths in step 4 are contained in $R_0$ (by choice of $s'$ and $t'$) and are shortest paths in $R_0$; and the subpath in step 5 is contained in the external separator $S$. It is also straightforward to see that the subpaths of $\cP_{k}$ form a partition of the edges of $P_{k}$. Finally, $\cP_{k}$ contains at most $7$ more subpaths than $\cP_{k-1}$ (+2 from step 2, +2 from step 3, +2 from step 4, and +1 from step 5). By induction, we conclude that $|\cP_{k}| \le 7k + 3$.
\end{proof}

\subsection{[Stretch] property.} 
\label{SSS:stretch}

We want to show that $\norm{P} \le (1+O(\log n \cdot \log \spread) \cdot \e)) \cdot \dist_{R_0^{\arcto}}(a, b)$. If every external separator of $R_0$ was a shortest path \emph{with respect to the entire graph} $G$, then proving the stretch property would be easy --- one could show that path $P_{k}$ has length at most $+\e \cdot \dist_{R^{\arcto}}(a,b)$
more than $P_{k-1}$ (as $P_{k}$ is constructed by ``detouring'' path $P_{k-1}$ through a nearby external separator, which is a shortest path). Indeed, in planar graphs we can construct a separator hierarchy where each separator is a shortest path with respect to $G$. However, in minor-free graphs, we can only guarantee that the separators we detour against are shortest paths in the subgraph $R_0$. As the paths $P_{k-1}$ are not entirely contained in $R_0$, we can no longer bound the length $P_k$ as any function of the length of $P_{k-1}$. Instead, we need a more involved argument, in which we strengthen the claim to enable induction --- rather than just bounding the length of $\norm{P_k}$, we bound the length of $\norm{P_k}$ even if we modify $P_k$ by replacing arbitrary subpaths of $P_k$ with shortest paths in $R_0$.

To this end, we introduce a new definition. 
For any path $P_k$ and subpath $X = P_k[x_1:x_2]$ of $P_k$,
we define the \EMPH{domain} of $X$, denoted \EMPH{$\domain(X)$}, to be the lowest region $R$ in the separator hierarchy such that (1) $R$ contains both \emph{endpoints} of $X$, and (2) $R$ is a (not necessarily proper) ancestor of $R_0$.
For any set of edge-disjoint subpaths $\cX$ of $P_k$, we define the \EMPH{domain replacement} of $P$ for $\cX$, denoted \EMPH{$\replace {P\/} {\cX}$}, to be the path obtained by starting with $P$ and replacing every subpath $X \in \cX$ with a shortest path in $\domain(X)$ between the endpoints of $X$. (Note that subpaths in $X$ may have endpoints that are not in $R_0$. This generality is not needed here, but it will be needed later in Section~\ref{S:proxy}.)

\begin{lemma}
\label{lem:safe-stretch}
    For every iteration $k \ge 0$, for any set of subpaths $\cX$ of $P_k$, we have 
    \[
    \norm {\replace {P_k} {\cX}} \le (1+ 4 k \e) \cdot \dist_{R_0^{\arcto}}(a,b).
    \]
\end{lemma}
\begin{proof}

    We proceed by induction on the iteration $k$. To simplify notation, we observe that $\norm{P_0} = \dist_{R_0^{\arcto}}(a, b)$ by construction, and prove that $ \norm{\replace {P_k} {\cX}} \le (1+4k\e) \cdot \norm{P_0}.$
    
    \bigskip \noindent In the \textbf{base case} ($k = 0$),
    we show that replacing any subpath $X = P_0[x_1:x_2]$ of $P_0$ with a shortest path in $\domain(X)$ does not increase the length of $P_0$. 
    Recall that $P_0$ is a shortest path in $R_0 \cup S_b$, where $S_b$ denotes the external separator of $R_0$ that contains $b$. Let $R_b$ be the region whose internal separator is $S_b$. There are two cases. (1) In the first case, either $x_1$ or $x_2$ lies on $S_b$.
    In this case, $\domain(X)$ includes $R_0 \cup S_b$, meaning that distances in $\domain(X)$ are no longer than distances in $R_0 \cup S_b$ --- thus swapping $X$ with $\spath_{\domain(X)}(x_1,x_2)$ does not increase the length of the path.
    (2) In the second case, both $x_1$ and $x_2$ are in $R_0$, so $\domain(X) = R_0$. Because shortest paths in $R_0 \cup S_b$ are unique, path $P_0$ is the concatenation of: a shortest path in $R_0$, a single edge between a vertex in $R_0$ and a vertex in $S_b$, and a subpath of $S_b$. By assumption, $X$ has both endpoints in $R_0$, so it is a subpath of a shortest path in $R_0$; thus swapping $X$ with $\spath_{\domain(X)}(x_1,x_2)$ has no effect on the length of the path.

    \bigskip \noindent In the \textbf{inductive case} ($k > 0$), we assume the claim is true for $P_{k-1}$. Let \EMPH{$S$} be the external separator considered during the construction of $P_k$, and let \EMPH{$s$}, \EMPH{$t$}, \EMPH{$s'$}, and \EMPH{$t'$} be the vertices chosen during \textsc{DetourPath} such that
    \[P_{k} = P_{k-1}[a:s] \circ \spath_{R_0 \cup S}(s, s') \circ \spath_{R_0 \cup S}(s',t') \circ \spath_{R_0 \cup S}(t', t) \circ P_{k-1}[t:b].\]
     
     We may assume without loss of generality that every subpath in $\cX$ is contained fully in $P_k[a:s]$, in $P_k[s:t]$, or in $P_k[t:b]$. Indeed, if some subpath $X = P_k[x_1:x_2]$ in $\cX$ contained vertices in $P_k[a:s]$ but was not fully contained in $P_k[a:s]$, then we could remove it from $\cX$ and add in two subpaths $X_1 = P_k[x_1:s]$ and $X_2 = P_k[s:x_2]$ to $\cX$ instead. Crucially, this swap only increases the length of $\replace{P_k}{\cX}$. This is because $s \in R_0$ and so $\domain(X)$ contains $\domain(X_1)$ and $\domain(X_2)$; thus, triangle inequality implies that
     \[
     \dist_{\domain(X)}(x_1, x_2) \le \dist_{\domain(X_1)}(x_1, s) + \dist_{\domain(X_2)}(s, x_2).
     \]
     A similar argument can be applied to subpaths that are not fully contained in $P_k[t:b]$. This completes the justification for our assumption, as we aim to prove an upper bound on the length of $\replace {P_k}{\cX}$.

    For any two vertices $u$ and $v$ on $P_k$, let \EMPH{$\cX_{u:v}$} denote the set of subpaths in $\cX$ that are entirely contained in $P_k[u:v]$. 
    Define $\EMPH{$\tilde{P}[u:v]$} \coloneqq \replace {P_k[u:v]} {\cX_{u:v}}$. By our assumption on $\cX$ from the previous paragraph, we have
    \begin{equation}
        \label{eq:swap-total}
        \replace {P_k} {\cX} = \tilde P[a:s] \circ \tilde{P}[s:t] \circ \tilde{P}[t:b].
    \end{equation}

    We first show that
    \begin{equation}
        \label{eq:swap-induction}
        \norm{\tilde{P}[a:s]} + \dist_{R_0}(s,t) + \norm{\tilde{P}[t:b]} \le (1 + 4(k-1)\e) \cdot \norm{P_0}.
    \end{equation}
    
    Indeed, define $\EMPH{$\cX'$} \coloneqq \cX_{a:s} \cup \set{P_{k-1}[s:t]} \cup \cX_{t:b}$, and notice that $\cX'$ is a set of subpaths of $P_{k-1}$ because $P_{k-1}[a:s]$ (resp.\ $P_{k-1}[t:b]$) is precisely the same path as $P_k[a:s]$ (resp.\ $P_k[t:b]$). Moreover, notice that the domain of the subpath $P_{k-1}[s:t]$ is $R_0$ (as $s,t \in R_0$). Thus the path $\replace {P_{k-1}}{\cX'}$ is precisely the same path as $\tilde{P}[a:s] \circ \spath_{R_0}(s,t) \circ \tilde{P}[t:b]$. By induction hypothesis, this path has length at most $(1 + 4(k-1))\e \cdot \norm{P_0}$.

    The rest of this proof is dedicated to showing that
    \begin{equation}
        \label{eq:swap-triangle}
        \norm{\tilde P[s:t]} \le \dist_{R_0}(s,t) + 4 \e \cdot \norm{P_0}.
    \end{equation}
    Together with Equations~\eqref{eq:swap-total} and \eqref{eq:swap-induction}, this will complete the proof of the inductive case.
    To prove Equation~\eqref{eq:swap-triangle}, we first observe that $P_k[s:t]$ (before replacements) has length at most $\dist_{R_0}(s, t) + 4 \e \norm{P_0}$. Indeed, 
     \begin{align*}
         \norm{P_k[s:t]} &= \norm{\spath_{R_0 \cup S}(s, s')} +
                            \norm{\spath_{R_0 \cup S}(s', t')} +
                            \norm{\spath_{R_0 \cup S}(t', t)}\\
                        &\le \e \norm{P_0} + \dist_{R_0 \cup S}(s', t') + \e \norm{P_0}\\
                        &\le \dist_{R_0 \cup S}(s, t) +4 \e \norm{P_0}
                        \\
                        &\le \dist_{R_0}(s, t) + 4 \e \norm{P_0}
     \end{align*}
     where the second-to-last inequality follows from triangle inequality.
     We now consider the effect (on the length of $P_k[s:t]$) of iteratively replacing each subpath $X$ in $\cX_{s:t}$ with a shortest path in $\domain(X)$ between the endpoints of $X$.
     After all subpaths $X \in \cX_{s:t}$ are swapped, we are left with the path $\tilde P[s:t]$.
     There are two cases.
     First observe that, as the subpaths of $\cX$ are edge-disjoint, there is at most one subpath $\EMPH{$\hat X$} = P_k[\hat x_1: \hat x_2]$ in $\cX_{s:t}$ such that its endpoints satisfy $\hat x_1 \in P_k[s:s']$, $\hat x_2 \in P_k[t':t]$, and $\hat x_1, \hat x_2 \in R_0$. 
     We argue below that starting with $P_k[s:t]$ and swapping $\hat X$ with $\spath_{\domain(\hat X)}(x_1, x_2)$ produces a path \EMPH{$\hat P$} with length at most $\dist_{R_0}(s,t) + 4\e \norm{P_0}$.
     (If no such subpath $\hat X$ exists, then we use $\hat P$ to denote $P_k[s:t]$, and we have $\norm{\hat P} \le \dist_{R_0}(s,t) + 4\e \norm{P_0}$ from our claim above.)
     Second, we argue that for any \emph{other} subpath $X \in \cX_{s:t}$, where $X \neq \hat X$, replacing subpath $X$ of $\hat P$ with a shortest path in $\domain(X)$ between its endpoints never increase the length of $\hat P$. As such, the path $\tilde P[s:t]$ that remains after all swaps have been completed has length at most $\dist_{R_0}(s,t) + 4 \e \norm{P_0}$.

     \bigskip \noindent \textbf{Case 1: Swapping out $\hat X$.}
     We claim that starting with $P_{k}[s:t]$ and swapping $\hat X$ for $\spath_{\domain(\hat X)}(s,t)$ produces a path $\hat P$ with length at most $\dist_{\domain(X)}(s,t) + 4 \e \norm{P_0}$.
     By assumption on $\hat x_1$ and $\hat x_2$, we have $\domain(\hat X) = R_0$.
     Thus $\hat P$ has length
     \[
     \norm{\hat P} = \norm{P_k[s:\hat x_1]} + \dist_{R_0}(\hat x_1, \hat x_2) + \norm{P_k[\hat x_2:t]}.
     \]
     By definition of $s$, $s'$, $t$ and $t'$, we have that $\norm{P_k[s:\hat x_1]} \le \e \norm{P_0}$ and $\norm{P_k[\hat x_2:t]} \le \e \norm{P_0}$. So $\norm{\hat P} \le 2 \e \norm{P_0} + \dist_{R_0}(\hat x_1, \hat x_2)$.
     Again by definition of $s$ and $t$, we have $\dist_{R_0}(s, \hat x_1) \le \e \norm{P_0}$ and $\dist_{R_0}(\hat x_2, t) \le \e \norm{P_0}$, so triangle inequality implies that $\dist_{R_0}(\hat x_1, \hat x_2) \le \dist_{R_0}(s,t)+ 2 \e \norm{P_0}$.
     We conclude that $\hat P$ has length at most $\dist_{R_0}(s,t) + 4 \e \norm{P_0}$.     

     \bigskip \noindent \textbf{Case 2: Swapping out $X \neq \hat X$.} 
     We claim that for every $X \in \cX_{s:t}$ with $X \neq \hat X$, we have $\norm{\spath_{\domain(X)} (x_1,x_2)} \le \norm X$, where $x_1$ and $x_2$ are the endpoints of $X$; that is, swapping out subpath $X$ does not increase the length.  Recall that $X$ is a subpath in $R_0 \cup S$. There are two subcases (from the fact that $X \neq \hat X$).     
     \begin{itemize}
     \item Suppose either $x_1$ or $x_2$ lies on the external separator $S$ considered during the construction of $P_k$. In this case $\domain(X)$ is the region whose internal separator is $S$, and so $R_0 \cup S$ is contained in $\domain(X)$.
     Thus $X$ is a path between $x_1$ and $x_2$ in $\domain(X)$. A shortest path between $x_1$ and $x_2$ in $\domain(X)$ can be no longer than $X$, so swapping out $X$ does not increase the length of $\hat P$.
     %
     \item Suppose that every vertex in $X$ is in $R_0$. In this case, $\domain(X) = R_0$. We have that $X$ is a subpath of either $P_k[s:s']$ or $P_k[t':t]$.
     But $P_k[s:s']$ and $P_k[t':t]$ are shortest paths in $R_0 \cup S$, so the subpath $X$ is a shortest path in $R_0 \cup S$.
     Because the subpath $X$ only contains vertices in $R_0$, it is a shortest path in $R_0$, and so swapping out $X$ does not change the length of $\hat P$.
     \end{itemize}
     This proves that $\norm{\tilde P[s:t]} \le \norm{\hat P} \le \dist_{R_0}(s,t) + 4\e \norm{P_0}$, as claimed in Equation~\eqref{eq:swap-triangle}.
\end{proof}

The [stretch property] follows from Lemma~\ref{lem:safe-stretch} and the observation that $\textsc{DetourPath}(\canon a b)$ returns a path $P$ within $O(\log n \cdot \log \spread)$ iterations because the outer loop runs $O(\log \spread)$ times, and the inner loop runs $O(\log n)$ times.

\begin{corollary}[Stretch property of $P$]
\label{cor:safe-stretch-simple}
    The length of $P$ is at most $(1 + O(\log n \cdot \log \spread) \cdot \e) \cdot \dist_{R^{\arcto}}(a,b)$. 
\end{corollary}

\subsection{[Bounded threats] property.}

The next section is devoted to proving the [bounded threats] property. Given a safe subpath $\pi$ of $P$, we want to bound the number of canonical pairs $\canon {a'} {b'}$ with $(\canon a b) \preceq (\canon {a'}{b'})$ that are threatened by $\pi$. Recall that $i_0$ and $R_0$ denote the scale and region (respectively) of $\canon a b$.
We will use Claim~\ref{clm:length-threat} in several places in this section, so we restate it here for convenience.
\lengththreat*

For the rest of this section, we fix $\pi$ to be an arbitrary safe subpath of $P$.
Rather than directly bounding the number of \emph{canonical pairs} threatened by $\pi$, we introduce the notion of a \emph{portal} being threatened, and show that it suffices to bound the number of portals threatened by $\pi$.
    A \EMPH{valid tuple $(i, R, p)$} is a tuple where $i \in [\log D]$ is a scale, $R$ is a region that is a (not necessarily proper) ancestor of $R_0$, and $p$ is an $(i, R)$-portal.
    We say that a valid tuple $(i, R, p)$ is \EMPH{threatened} by $\pi$ if
    \[
    \dist_{R}(\pi, p) \le 2 \cdot 2^i.
    \]
Rephrasing the definition of ``threatening a canonical pair'' from Definition~\ref{def:threatening}, we say an $(i, R)$-canonical pair $\canon {a'} {b'}$ is threatened by $\pi$ if the tuple $(i, R, a')$ is threatened by $\pi$.

\begin{claim}
\label{clm:portal-threat-reduction}
    Define variables $\kappa_{0}$ and $\kappa_{>0}$ as follows:
    \begin{itemize}
        \item \EMPH{$\kappa_0$} is the number of valid tuples $(i, R, p)$ threatened by $\pi$, where $R = R_0$ and $i \ge i_0$;
        \item \EMPH{$\kappa_{>0}$} is the number of valid tuples $(i, R, p)$ threatened by $\pi$, where $R$ is a proper ancestor of $R_0$.
    \end{itemize}
    There are at most $O(\e^{-1} \log n \cdot (\kappa_0 + \kappa_{>0}))$ canonical pairs $\canon {a'} {b'}$ threatened by $\pi$ where $(\canon a b) \preceq (\canon {a'}{b'})$.
\end{claim}

\begin{proof}
For every $(i, R)$-canonical pair $\canon {a'} {b'}$ threatened by $\pi$ where $(\canon a b) \preceq (\canon {a'}{b'})$, we add a \EMPH{charge} to the valid tuple $(i, R, a')$. 
By definition of threatening, $(i, R, a')$ is threatened by $\pi$. By definition of $\preceq$, either $R = R_0$ and $i \ge i_0$, or $R$ is a strict ancestor of $R_0$. Thus, the total number of valid tuples $(i, R, a')$ that receive a charge is at most $\kappa_0 + \kappa_{>0}$.

To complete the proof, we let $(i, R, a')$ be an arbitrary valid tuple and show that it is charged at most $O(\e^{-1} \log n)$ times. Indeed, this tuple is charged once for every $b'$ such that $\canon{a'}{b'}$ is an $(i, R)$-canonical pair. By definition of canonical pair, this means that the tuple is charged once for every $b'$ such that $b'$ is an $(i, R')$-portal for some region $R'$ which is an ancestor of $R$, where $\dist_{R'}(a', b') \le 2^i$. But there are only $O(\log n)$ ancestor regions $R'$ of $R$, and Claim~\ref{clm:length-threat} implies that for each one, there are only $O(\e^{-1})$ many $(i, R')$-portals $b'$ with $\dist_{R'}(a',b')\le 2^i$. Thus $(i, R, a')$ is charged only $O(\e^{-1} \log n)$ times.
\end{proof}

Bounding $\kappa_0$ is straightforward.

\begin{claim}
\label{clm:kappa-0}
    There are $O(\e^{-1} \log \spread)$ valid tuples $(i, R, p)$ that are threatened by $\pi$ such that $R = R_0$ and $i \ge i_0$. In other words, $\kappa_0 = O(\e^{-1} \log \spread)$.
\end{claim}
\begin{proof}
    There are $O(\log \spread)$ scales $i \in [\log \spread]$  such that $i \ge i_0$. For each scale, Claim~\ref{clm:length-threat} implies that there are only $O(2 \cdot \e^{-1} \cdot \ceil{ \frac{\norm{\pi}}{2^i}}) = O(\e^{-1} \cdot \ceil{ \frac{2^{i_0}}{2^i}}) = O(\e^{-1})$ portals $p$ such that $(i, R_0, p)$ is threatened by~$\pi$; the first equality follows from the fact that $\norm{\pi} \le O(2^{i_0})$ by \Cref{cor:safe-stretch-simple} and the fact that $\e < \frac{1}{O(\log^3 n \cdot \log \spread)}$.
\end{proof}

Bounding $\kappa_{>0}$ is more delicate, as we may not assume $i \ge i_0$. We first introduce a stronger notion of threatening a portal, and bound the number of strongly-threatened portals (Claim~\ref{clm:interior-threat}). We then use a charging argument to bound the number of threatened portals in terms of the number of strongly-threatened portals (Claim~\ref{clm:external-threats}).
For any $\alpha \ge 1$ and valid tuple $(i, R, p)$, we say that $(i, R, p)$ is \EMPH{$\alpha$-strongly-threatened} by $\pi$ if
    \[
    \dist_{R_0 \cup S}(\pi, p) \le \alpha \cdot 2^i.
    \]
    where $S$ denotes the internal separator of $R$. 
The difference between the definition of \emph{threatened} and \emph{strongly-threatened} is the subgraph in which distance is measured. 
As $R_0 \cup S$ is a subgraph of $R$ (recall that $R$ is an ancestor of $R_0$), any $(i, R', p)$ tuple that is $2$-strongly-threatened by $\pi$ is also threatened  (normally) by $\pi$.

\begin{figure}
    \centering
    \includegraphics[width=0.6\linewidth]{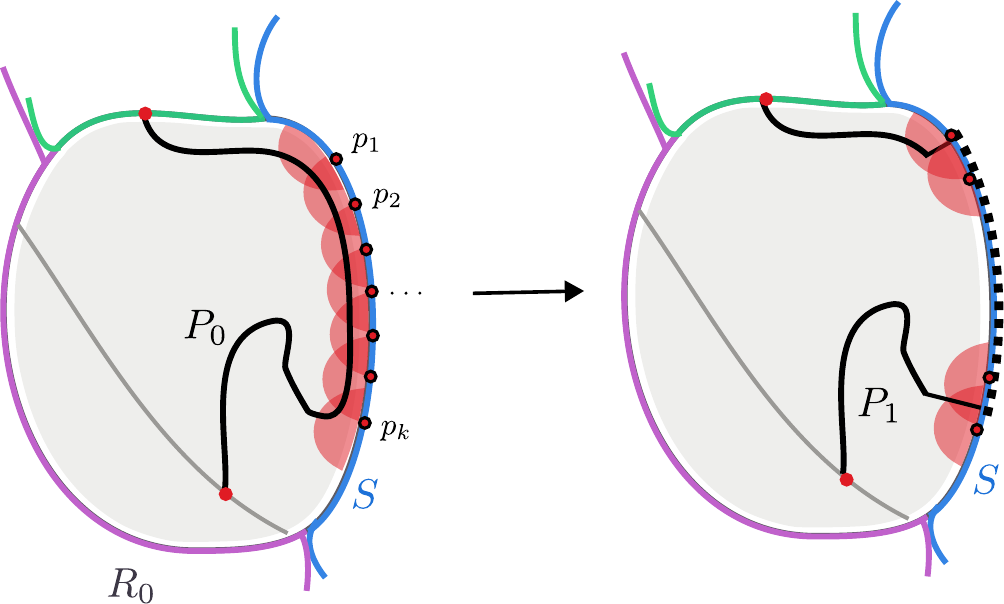}
    \caption{Intuition for the proof of \Cref{clm:interior-threat}. On the left: A region $R_0$ and a path $P_0$ which is $\alpha$-strongly threatened by the tuples $(i, R, p_1)$, $(i, R, p_2)$, ..., $(i, R, p_k)$; here $R$ denotes the region (not drawn) whose internal separator is $S$ (drawn in blue). The red neighborhoods around the portals $p_j \in \set{p_1, \ldots, p_k}$ represent an $\alpha \cdot 2^i$ neighborhood around $p_j$ in the graph $R_0 \cup S$. On the right: After $P_9$ is detoured against $S$ to produce $P_1$, the safe subpaths of $P_1$ are threatened by fewer tuples $(i, R, p_j)$.}
    \label{fig:threatening}
\end{figure}

\begin{claim}
\label{clm:interior-threat}
    There are $O(\e^{-2} \log \spread \log n)$ valid tuples $(i, R, p)$ that are 4-strongly-threatened by $\pi$, such that region $R$ is a proper ancestor of $R_0$.
\end{claim}

\begin{proof}
There are $O(\log n)$ ancestor regions of $R_0$ and $O(\log \spread)$ scales. For the rest of the proof, we fix a scale $i \in [\log \spread]$ and ancestor region $R$, and prove that there are $O(\e^{-2})$ portals $p$ with scale $i$ and region $R$ such that $(i, R, p)$ is $4$-strongly-threatened by $\pi$.

Suppose that $i > \lceil \log_2 (\e \norm{P_0}) \rceil - 2$. As $\norm{\pi} = O(\norm{P_0})$ and $2^i = \Omega(\e \norm{P_0})$, Claim~\ref{clm:length-threat} implies that there are only $O(\e^{-1} \cdot \ceil{\frac{\norm{\pi}}{2^i}}) = O(\e^{-2})$ many $(i, R)$-portals $p$ such that $\dist_{R}(\pi, p) \le 4 \cdot 2^i$ (which is a condition that certainly holds if $\dist_{R_0 \cup S}(\pi, p) \le 4 \cdot 2^i$), and the claim holds.
Otherwise, we have $i \in [\log_2 (\e \norm{P_0}) - 2]$, and so there is some iteration of the outer loop of \textsc{DetourPath} that processes scale $i + 2$. Let $S$ be the internal separator of $R$. Let \EMPH{$K$} denote the value of the iteration counter $k$ during the execution of $\textsc{DetourPath}$ at the time when scale $i + 2$ and external separator $S$ is processed; that is, path $P_K$ is created by detouring $P_{K-1}$ against $S$ if $P_{K-1}$ is within distance $4 \cdot 2^{i}$ of $S$. 
We prove:
\begin{quote}
    For every $k \ge K$ and a subpath $\hat \pi$ of a safe subpath of $P_k$\footnote{we emphasize $\hat \pi$ is not an arbitrary path; rather, it is (a subpath of) some safe subpath of $P_k$ from the partition $\cP_k$ defined in Lemma~\ref{lem:detour-subpath}},
    there are $O(\e^{-1})$ many $(i, R)$-portals $p$ such that $\dist_{R_0 \cup S}(\hat \pi, p) \le 4 \cdot 2^i$.
\end{quote}
This suffices to prove the claim, as $\pi$ is a safe subpath of the final path $P_k$ returned by $\textsc{DetourPath}$.
We proceed by induction on $k$. Refer to \Cref{fig:threatening}.

\medskip \noindent
\emph{As a base case}, we consider iteration $K$. Let $\hat \pi$ be a subpath of a safe subpath of $P_K$ (as defined in the proof of \Cref{lem:detour-subpath}); in particular, $\hat \pi$ is a shortest path in $R_0$. 
Let $s$, $s'$, $t'$, and $t$ be the vertices defined in the $K$ iteration of $\textsc{DetourPath}$; that is, $P_{K}$ is the concatenation $P_{K-1}[a:s] \circ \spath_{R_0 \cup S}(s, s') \circ \spath_{R_0 \cup S}(s', t') \circ \spath_{R_0 \cup S}(t', t) \circ P_{K-1}[t:b]$. 
\begin{itemize}
    \item If $\hat \pi$ is a subpath of $P_{K-1}[a:s]$, then (by choice of $s$), the only vertex of $\hat \pi$ within distance $4 \cdot 2^i$ of $S$ in $R_0 \cup S$ is $s$; thus, the claim holds. Similarly, if $\pi$ is a subpath of $P_{K}[t:b]$, the claim holds by choice of $t$.  
    \item Otherwise, $\hat \pi$ must be a subpath of either $\spath_{R_0 \cup S}(s, s')$ or $\spath_{R_0 \cup S}(t', t)$. (Note that $\spath_{R_0 \cup S}(s', t')$ is disjoint from $R_0$, while $\hat \pi$ is in $R_0$ by assumption.) In either case, $\norm{\hat \pi} \le 4 \cdot 2^i$, and so Claim~\ref{clm:length-threat} implies that there only $O(\e^{-1})$ many $(i, R)$-portals with $\dist_{R}(\hat\pi, p) \le 4\cdot 2^i$. 
9    The claim now follows from the fact that $R_0 \cup S$ is a subgraph of $R$, so the number of portals $p$ satisfying $\dist_{R_0 \cup S}(\hat\pi, p) \le 4\cdot 2^i$ is at most those satisfying $\dist_{R}(\hat\pi, p) \le 4\cdot 2^i$.
\end{itemize}

\noindent
\emph{The inductive case} is similar. We consider iteration $k > K$. 
As above, let $\hat \pi$ be a subpath of a safe subpath of $P_k$ (as defined in the proof of \Cref{lem:detour-subpath}). Let $s$, $s'$, $t'$, and $t$ be the vertices used to define $P_k$ during the execution of \textsc{DetourPath}. Again, there are two cases.
\begin{itemize}
    \item If $\hat \pi$ is a subpath of $P_{k-1}[a:s]$ (resp.\ of $P_{k-1}[t:b]$), then (by the proof of \Cref{lem:detour-subpath}) it is (a subpath of) some safe subpath of $P_{k-1}$.
    The claim follows from the 0-inductive hypothesis.
    \item Otherwise $\hat \pi$ must be a subpath of either $\spath_{R_0 \cup S}(s, s')$ or $\spath_{R_0 \cup S}(t', t)$. As $k > K$, the current iteration considers a scale upper-bounded by $i+2$; thus, $\norm{\hat \pi} \le 4 \cdot 2^i$ using an argument similar to the base case, and the claim follows from Claim~\ref{clm:length-threat}. 
    \qed
\end{itemize}
\vspace{-22pt}
\end{proof}

\begin{claim}
\label{clm:external-threats}
    There are $O(\e^{-3} \log^2 n \log \spread)$ valid tuples $(i, R, p)$ that are threatened by $\pi$, such that region $R$ is a proper ancestor of $R_0$. In other words, $\kappa_{>0} = O(\e^{-3} \log^2 n \log \spread)$.
\end{claim}

\begin{proof}
Let $(i, R, p)$ be a tuple threatened by $\pi$. Let $S$ be the internal separator of $R$, so that $p$ lies in $S$. 
As $R$ is a proper ancestor of $R_0$, the separator $S$ is not contained in $R_0$, and in particular $p \not \in R_0$. 
Consider some shortest path $P_p$ in $R$ from $p$ to $\pi$ (which has length at most $2 \cdot 2^i$), and let \EMPH{$x$} be the last vertex along $P_p$ that is \emph{not} contained in $R_0$ (note that such a vertex always exists, as $p \not \in R_0$). 
By definition of separator hierarchy, $x$ is on some external separator $S_x$ of $R_0$. 
We have $\dist_{R_0 \cup S_x}(\pi, x) \le \norm{P_p} \le 2 \cdot 2^i$.

Let \EMPH{$R_x$} be the region whose internal separator is $S_x$. By definition of scale-$i$ portals, there is some $(i, R_x)$-portal \EMPH{$x'$} on $S_x$ with $\dist_{S_x}(x, x') \le \e\cdot 2^i$. 
By triangle inequality and $\e < 1$, $\dist_{R_0 \cup S_x}(\pi, x') \le 4 \cdot 2^i$. 
We \EMPH{charge} the threatened tuple $(i, R, p)$ to the 4-strongly-threatened tuple $(i, R_x, x')$, and observe that $R_x$ is a proper ancestor of $R_0$.

By Claim~\ref{clm:interior-threat}, there are $O(\e^{-2} \log \spread \log n)$ tuples of the form $(i, R_x, x')$ that are 4-strongly-threatened by $\pi$ and have $R_x$ as a proper ancestor of $R_0$. We now argue that each such tuple $(i, R_x, x')$ is charged at most $O(\e^{-1} \log n)$ times. Indeed, the tuple $(i, R_x, x')$ is only charged by some other tuple $(i, R, p)$ where (1) $R$ is an ancestor of $R_x$, (2)  $p$ is an $(i, R)$-portal, and (3) $\dist_{R}(p, x') \le 2 \cdot 2^i$. There are $O(\log n)$ ancestor regions $R$ of $R_x$, and \Cref{clm:length-threat} implies that (for each $R$) there are only $O(\e^{-1})$ scale-$i$ portals within distance $2 \cdot 2^i$ of $x'$.
This proves the claim.
\end{proof}

\begin{lemma}[Bounded threats property for $P$]
    There are 
    \(
    {O}(\e^{-4} \log^3 n \log \spread)
    \)
    canonical pairs $\canon {a'} {b'}$ with $(\canon a b) \preceq (\canon {a'}{b'})$ that are threatened by $\pi$
    for any given safe subpath $\pi$ of $P$.
\end{lemma}
\begin{proof}
This follows immediately from Claims~\ref{clm:portal-threat-reduction}, \ref{clm:kappa-0}, and \ref{clm:external-threats}.
\end{proof}

\subsection{[Bounded splits] property.}

Let $\pi$ be a safe subpath of $P$. 
This section is dedicated to proving the [bounded splits] property. That is, given a shortest path $\pi_{\rm ext}$ in some ancestor region $R_{\rm ext}$ of $R_0$, we want to show that there are few splitting points between $\pi$ and $\pi_{\rm ext}$. 
We remark that if each external separator was a shortest path in $G$, we could show that $\pi_{\rm ext}$ enters and exits the region $R_0$ only $\tilde O_\e(1)$ times; the fact that both $\pi$ and $\pi_{\rm ext}$ are shortest paths would bound the number of splitting points each time $\pi_{\rm ext}$ entered. However, in minor-free graphs, external separators do not have this property --- $\pi_{\rm ext}$ could enter and exit $R_0$ an unlimited number of times. Rather than just relying on the fact that $\pi$ is a shortest path in $R_0$, we need a more subtle argument that exploits the fact $\pi$ was constructed during \textsc{DetourPath}.
We show in Lemma~\ref{lem:bounded-splits} that in such case $\pi$ can only enter and exit $R_0$ for $\tilde{O}(\e^{-1})$ many times.

\begin{claim}
\label{clm:two-points-close}
    Let $S$ be an external separator of $R_0$. 
    If the safe subpath $\pi$ contains more than one edge, then there is at most one point $x$ on $\pi$ that satisfies $\dist_{R_0 \cup S}(x, S) \le \e \norm{\pi}/2$.
\end{claim}

\begin{proof}
Assume that $\e \norm{\pi}/2 \ge 1$; otherwise, there are no points $x$ with $\dist_{R_0 \cup S}(x, S) \le \e \norm{\pi}/2$ because every edge has length at least $1$.
Let $i \in [\log \spread]$ be the smallest scale satisfying $2^i \ge \e \norm{\pi}/2$; that is,
\[
\EMPH{$i$} \coloneqq \lceil \log_2 ( \e \norm{\pi}/2) \rceil.
\]
Notice that $i \ge 0$ because $\e \norm{\pi}/2 \ge 1$.
Further, $i \le \lceil \log (\e \norm{P_0}) \rceil$ (recall that $P_0$ is the shortest path between the input canonical pair to \textsc{DetourPath}); this is because $\norm \pi \le (1+4\log n \log\spread \cdot \e)\norm{P_0} < 2\norm{P_0}$, by Lemma~\ref{lem:safe-stretch} and the fact that $\e = \frac{\e_0}{\Theta(\log^3 n \cdot \log \spread)} < \frac{1}{4 \log n \cdot \log \spread}$.
The outer for-loop of \textsc{DetourPath} processes every scale between $0$ and $\lceil \log (\e \norm{P_0}) \rceil$, and in particular some iteration of the for-loop processes scale $i$.
Let $\EMPH{$K$} \ge 1$ denote the iteration number during which scale $i$ and external separator $S$ are processed (i.e., the iteration during which path $P_{K}$ is created by detouring $P_{K-1}$ along $S$).
    
We first show that $\pi$ is a subpath of some safe subpath of $\cP_K$, the subpath decomposition of $P_K$ as defined in Lemma~\ref{lem:detour-subpath}. 
Indeed, an inductive argument shows that for every future iteration $k > K$, every subpath in $\cP_{k}$ is either: (1) a subpath (of a subpath) in $\cP_{K}$, (2) a single edge,  (3) contained in an external separator, or (4) of length $\le 2^i$. 
By assumption, $\pi$ is a safe subpath with more than one edge, so $\pi$ does not fall into cases (2) or (3). Moreover, we claim that $\pi$ does not fall into case (4) because $\norm{\pi} > 2^i$. 
Indeed,
\[
2^i \le 2^{\log_2 (\e \norm \pi /2) + 1} = \e \norm \pi < \norm \pi.
\]

Thus $\pi$ falls into case (1) --- that is, $\pi$ is (a subpath of) a subpath in $\cP_{K}$. 
Suppose for the sake of contradiction (to our main claim) that there are two points $x_1$ and $x_2$ on $\pi$ that are within distance $\e \norm{\pi}/2 \le 2^i$ of $S$ (where distance is measured in the graph $R_0 \cup S$). Recall that vertices $s$ and $t$ in iteration $K$ of \textsc{DetourPath} are the first and last vertices on $P_{K-1}$
within distance $2^i$ of $S$; thus, $x_1, x_2 \in P_{K}[s:t]$, and so $\pi$ is a subpath in $\cP_{K}$ that lies in $P_{K}[s:t]$. 
However, (by construction) every such subpath in $\cP_{K}$ that lies in $P_{K}[s:t]$ either has length at most $2^i$, is a single edge, or is contained in $S$. The subpath $\pi$ fits into none of these three categories, a contradiction.
\end{proof}

\begin{observation}
\label{obs:two-splits}
    Let $\pi_{\rm ext}$ be a shortest path in some region $R_{\rm ext}$ that is an ancestor of $R_0$.
    Let $p_1$ and $p_2$ be points on $\pi_{\rm ext}$ that are on external separators of $R_0$, such that every vertex of $\pi_{\rm ext}[p_1:p_2]$ other than $p_1$ and $p_2$ is contained in $R_0$.\footnote{$p_1$ and $p_2$ may or may not be in $R_0$.}
    Then there are at most $2$ splitting points of $\pi \cap \pi_{\rm ext}[p_1:p_2]$.
\end{observation}
\begin{proof}
    This follows from the fact that $\pi_{\rm ext}[p_1:p_2]$ (excluding the endpoints) is a shortest path in $R_0$, and shortest paths in $R_0$ are unique.
\end{proof}

\begin{lemma}[Bounded splits property for $P$]
\label{lem:bounded-splits}
    Let $\pi_{\rm ext}$ be a shortest path in some region $R_{\rm ext}$ that is an ancestor of $R_0$. Then there are $O(\e^{-1}\log n)$ splitting points of $\pi \cap \pi_{\rm ext}$.
\end{lemma}

\begin{proof}
We first claim that WLOG we may assume $\norm{\pi_{\rm ext}} \le \norm{\pi}$. Indeed, let $p_1$ and $p_2$ be the first and last intersection points of $\pi$ and $\pi_{\rm ext}$, as one travels along $\pi_{\rm ext}$ in an arbitrary direction. Every splitting point of $\pi \cap \pi_{\rm ext}$ is also a splitting point of $\pi \cap \pi_{\rm ext}[p_1:p_2]$, except for possibly $p_1$ and $p_2$. 
As we aim to prove a bound of $O(\e^{-1}\log n)$ splitting points, it suffices to bound the splitting points of $\pi_{\rm ext}[p_1:p_2]$. 
But observe that $\norm{\pi_{\rm ext}[p_1:p_2]} \le \norm{\pi[p_1:p_2]} \le \norm{\pi}$, as $\pi_{\rm ext}[p_1:p_2]$ is a shortest path between $p_1$ and $p_2$ in $R_{\rm ext}$ (which is a region that contains $R$ and thus the path $\pi$). 
For the rest of the proof, we assume that $\norm{\pi_{\rm ext}} \le \norm{\pi}$.
We also assume WLOG that $\pi$ consists of more than a single edge (as otherwise there are trivially at most 2 splitting points).

Imagine chopping $\pi_{\rm ext}$ into edge-disjoint subpaths by cutting at every vertex that is not in $R_0$. Many of these resulting subpaths consist of a single edge outside of $R_0$ (and thus do not intersect $\pi$); the other subpaths have all vertices except their endpoints contained in $R_0$ (and could intersect $\pi$). 
Let
\(
\EMPH{$\Pi$} = \set{\pi_i, \ldots, \pi_\beta}
\)
denote the set of subpaths of $\pi_{\rm ext}$ (after cutting as described above) that contain at least one vertex in $\pi$. Every splitting point $x$ of $\pi \cap \pi_{\rm ext}$ is in $R_0$ and is therefore not an endpoint of any subpath of $\pi_{\rm ext}$; therefore, there is some subpath $\pi_i \in \Pi$ such that $x$ is a splitting point of $\pi \cap \pi_i$.
By Observation~\ref{obs:two-splits}, for each subpath $\pi_i \in \Pi$ there are at most 2 splitting points of $\pi \cap \pi_i \in \Pi$. We now show that $|\Pi|$ is at most $O(\e^{-1}\log n)$.

If $|\Pi| = 1$, then the claim follows trivially. Otherwise, every subpath $\pi_i$ of $\Pi$ has at least one endpoint that touches some external separator $S_i$ of $R_0$.  We \EMPH{charge} $\pi_i$ to separator $S_i$. We show that each of the $O(\log n)$ external separators is charged $O(\e^{-1})$ times.
To this end, let \EMPH{$\Delta$} denote the smallest distance such that there exists an external separator $S$ and two points $p_1$, $p_2$ on $\pi$ with $\dist_{R_0 \cup S}(p_1, S) \le \Delta$ and $\dist_{R_0 \cup S}(p_2, S) \le \Delta$. By Claim~\ref{clm:two-points-close}, we have $\norm{\pi} \le O(\Delta/\e)$ and in particular
\(
\norm{\pi_{\rm ext}} \le O(\Delta/\e). 
\)
If a subpath $\pi_i$ charges separator $S_i$, there is a vertex $x_i$ on $\pi$ (namely, a splitting point of $\pi \cap \pi_i$) such that $\dist_{R_0 \cup S_i}(x_i, S_i) \le \norm{\pi_i}$; this is because $\pi_i$ is a path in $R_0 \cup S_i$ that contains both $x_i$ and a vertex on $S_i$. If subpath $\pi_i$ charges a separator $S_i$ that has been charged before, then definition of $\Delta$ implies that
\(
\norm{\pi_i} \ge \Delta.
\)
As $\sum_{i}\norm{\pi_i} \le \norm{\pi_{\rm ext}} \le O(\Delta/\e)$, we conclude that there are at most $O(1/\e)$ subpaths that charge separators which have been charged before. There are $O(\log n)$ external separators (each of which can be ``safely'' charged once), and so $|\Pi| = O(\e^{-1} \log n)$. \qed
\end{proof}

\section{Proof of Lemma~\ref{lem:rel-pairs}:  Relevant pairs of a terminal}
\label{S:rel-pairs}

This section is dedicated to proving \Cref{lem:rel-pairs}. For any vertex $v$ in $V(G)$, we define a set \EMPH{$X_v$} of \EMPH{relevant portals} as follows: for every region $R$ in the separator hierarchy that contains $v$, for every scale $i \in [\lceil \log \spread \rceil]$, for every $(i, R)$-portal $p$, take $p$ as a relevant portal if $\dist_R(p, v) \le 2^i$.
We define the \EMPH{relevant pairs} for $v$, denoted \EMPH{$\operatorname{RelPairs}(v)$}, to be the set of all canonical pairs $\canon a b$ with $a, b \in X_v$.

\begin{claim}
    For every vertex $v \in V(G)$, the set $\operatorname{RelPairs}(v)$ contains $O(\e^{-2} \cdot \log^2 n \cdot \log^2 \spread)$ pairs.
\end{claim}
\begin{proof}
    There are $O(\log n)$ regions $R$ containing $v$ and there are $O(\log \spread)$ scales $i$. By \Cref{clm:length-threat}, the number of $(i, R)$-portals $p$ with $\dist_R(v, p) \le 2^i$ is $O(\e^{-1})$. Thus $|X_v| = O(\e^{-1} \cdot \log n \cdot \log \spread)$. The number of relevant pairs of $v$ is at most $|X_v|^2$.
\end{proof}

Given vertices $u$ and $v$, we want to find a canonical sequence between $u$ and $v$ using only $\operatorname{RelPairs}(u) \cup \operatorname{RelPairs}(v)$. We begin with a special case.
\begin{claim}
\label{clm:rel-helper}
    Let $a$ be a vertex in $V(G)$, which lies on the internal separator $S$ of some region $R$. Let $p$ be a vertex on $S$ which is in the relevant vertex set $X_a$. 
    There is an (exact) $1$-canonical sequence $(x_1, \ldots, x_\ell)$ between $a$ and $p$ with respect to the subgraph $G[R]$, such that every $\canon{x_i}{x_{i+1}}$ is a canonical pair in $\operatorname{RelPairs}(a)$. Moreover, every vertex $x_i$ in the sequence lies on $S$, and $\ell = O(\log \spread)$.%
    \footnote{The property of the last sentence will not be used in this section, but it helps later in Section~\ref{S:proxy}.}
\end{claim}

\begin{proof}    
    Observe that the shortest path between $a$ and $p$ in $R$ is along $S$. For any vertex $x \in S$, we define \EMPH{$\tau(x)$} to be the \EMPH{largest scale} such that $x$ is a $(\tau(x), R)$-portal.
    Note that every $x \in S$ is a $(0, R)$-portal, so $\tau(x)$ is well-defined. Also observe that $x$ is a scale-$t$ portal for every $t \le \tau(x)$.
    
    We now iteratively define a sequence of vertices \EMPH{$(x_1, \ldots, x_k)$} on $S$ as follows. Define $x_1$ to be $a$. For every $i > 1$, we first check if $\tau(x_{i-1}) \ge \tau(p)$; if so, take $\EMPH{$x_i$} \coloneqq p$ and terminate. Otherwise, scan over the vertices of $S$ in order from $x_{i-1}$ to $p$, and we let \EMPH{$x_i$} be the first vertex with $\tau(x_i) \ge 
    \tau(x_{i+1}) + 1$.
    As there are only $O(\log \spread)$ scales, this procedure terminates with $k = O(\log \spread)$.
    
    We now have to show that each $\canon {x_{i-1}}{x_i}$ is a well-defined canonical pair, and that it is in $\operatorname{RelPairs}(a)$.
    We first consider the case where $x_i \neq p$. We show that $x_i$ is in $X_a$. 
    Inductively maintain the invariant that $\dist_S(a, x_i) \le 2^{\tau(x_i)}$: This clearly holds in the base case when $x_1 = a$.
    In the inductive case, we have $x_{i-1}$ is within distance $2^{\tau(x_{i-1})} \le 2^{\tau(x_i)}/2$ of $a$. The definition of portals means that we run into the scale-$\tau(x_i)$ portal $x_i$ after walking distance at most $\e 2^{\tau(x_i)}$ along $S$ from $x_{i-1}$. The total distance from $a$ to $x_i$ is $\e 2^{\tau(x_i)} + 2^{\tau(x_i)}/2 \le 2^{\tau(x_i)}$. This implies that $x_i$ is in $X_a$, as required. Moreover, a similar argument implies $\dist_R(x_{i-1}, x_i) \le \e 2^{\tau(x_{i-1}) + 1} \le 2^{\tau(x_{i-1})}$. This means that $\canon {x_{i-1}} {x_i}$ is in fact a canonical pair, with scale $\tau(x_{i-1})$. Thus, $\canon {x_{i-1}} {x_i}$ is in $\operatorname{RelPairs}(a)$.

    Finally, consider the last pair $\canon{x_{k-1}} {x_k}$, where $x_k = p$. As $p$ is in $X_a$, we have $\dist_R(a, p) \le 2^{\tau(p)}$. As $x_{k-1}$ can only be closer to $p$ than $a$, we have $\dist_R(x_{k-1}, x_k) \le 2^{\tau(p)} = 2^{\tau(x_{k-1})} $. This means that $\canon {x_{k-1}} {x_k} $ is indeed a canonical pair (at scale $\tau(p)$), and thus it is in $\operatorname{RelPairs}(a)$.
\end{proof}

We are now ready to find a canonical sequence for the general case.
\begin{proof}[of \Cref{lem:rel-pairs}]
    Let \EMPH{$P$} be the shortest path in $G$ between $a$ and $b$. Let \EMPH{$R$} be the lowest region in the separator hierarchy that contains $P$. It follows from the definition of separator hierarchy (specifically, the property that the children of $R$ are connected components of $G[R \setminus S]$) that $P$ intersects the internal separator $S$ of $R$. 
    Let \EMPH{$x$} denote one such intersection point in $S \cap P$.
    Let \EMPH{$s$} be the scale such that $\norm{P} \in [2^{s-2}, 2^{s-1})$. 
    By definition of portals, there is an $(s, R)$-portal $p$ within distance $\e \norm{P}$ of $x$; that is, $\dist_R(x, p) \le \e 2^{s-2} = O(\e \norm{P})$. By triangle inequality, we have $\dist_R(a, p) \le \dist_R(a, x) + \dist_R(x, p) \le (1+O(\e)) \cdot \norm{P} \le 2^s$. Similarly $\dist_R(b, p) \le 2^s$. Therefore, portal $p$ is in the \emph{relevant vertex sets} $X_a$ and $X_b$ for both $a$ and $b$, respectively. 
    In the remainder of the proof, we will find a $(1+O(\log n)\cdot \e)$-canonical sequence from $a$ to the portal $p$, using only canonical pairs in $\operatorname{RelPairs}(a)$. A symmetric argument implies that one can find a $(1+O(\log n)\cdot \e)$-canonical sequence from $b$ to the portal $p$ using only canonical pairs in $\operatorname{RelPairs}(b)$. Concatenating these two canonical sequences proves the lemma.

    Let \EMPH{$R_1$} be the region such that $a$ lies on the internal separator of $R_1$. Define \EMPH{$x_1$} to be the $(s, R_1)$-portal closest to $a$. (Recall that $s$ is the scale of portal $p$). Clearly $x_1$ is in $\operatorname{RelPairs}(a)$, as $\dist_{R_1}(s, R_1) \le \e 2^s \le 2^s$.
    By \Cref{clm:rel-helper}, there is a 1-canonical sequence from $p$ to $x_1$.
    Now consider the shortest path \EMPH{$\tilde P$} in $R_1$ from $x_1$ to portal $p$. 
    We imagine orienting $\tilde P$ so that it starts at $x_1$ and ends at $p$. 
    We inductively define a sequence of vertices \EMPH{$(x_1', x_2', \ldots, x_k')$} that lie on $\tilde P$, and a sequence of vertices \EMPH{$(x_1, x_2, \ldots, x_k)$} which are scale-$s$ portals.
    We set $\EMPH{$x'_1$} \gets x_1$.
    For every $i > 1$, define \EMPH{$x'_i$} to be the first vertex along $\tilde P$ that leaves the region $R_{i-1}$; let \EMPH{$R_i$} be the region whose internal separator contains $x'_i$; and let \EMPH{$x_i$} be the closest $(s,R_i)$-portal to $x'_i$ (meaning that $x_i$ lies on the same internal separator that contains $x'_i$).
    
    \medskip \noindent \textbf{Size.~} First observe that $R_i$ is a proper ancestor of $R_{i-1}$, so this process terminates after $O(\log n)$ iterations, when we pick $x_k \gets p$.  (In particular, $k = O(\log n)$.)
    
    \medskip \noindent \textbf{Canonical sequence.~} Next observe that $\canon {x_i} {x_{i+1}}$ is a canonical pair; in particular, it is an $(s, R_i)$-canonical pair, as both $x_i$ and $x_{i+1}$ are scale-$i$ portals. Moreover, each $x_i$ is in the set $X_a$ of relevant vertices for $a$: indeed, triangle inequality implies that
    \[\dist_{R_i}(x_i, a) \le \dist_{R_i}(x_i, x_i') + \dist_{R_i}(x_i', a) \le \e 2^s + \norm{P}.\]
    The last inequality follows from the fact that $x'_i$ lies on the path $\tilde P$, and the subpath of $\tilde P$ running from $a$ to $x'_i$ is entirely contained in $R_i$. Finally, $\norm{\tilde P} \le (1+O(\e)) \cdot \norm{P} \le (1+O(\e)) \cdot 2^{s-1}$, so we have $\dist_{R_i}(x_i, a) \le 2^s$, meaning that $x_i$ is added to the set $X_a$. Similarly, $x_{i+1}$ is in $X_a$. Thus $\canon{x_i} {x_{i+1}}$ is in $\operatorname{RelPairs}(a)$.

    \medskip \noindent \textbf{Stretch.~} It remains to bound the stretch 
    \[\sum_{i=1}^{k - 1} \dist_{R_i^{\arcto}}(x_i, x_{i+1}).\]
    By definition of portals, the vertex $x_i'$ is within distance $\e 2^s$ of $x_i$, where distance is measured with respect to the internal separator of $R_i$.
    This implies that $\dist_{R_i^{\arcto}}(x_i, x_i') \le \e 2^s$. 
    Similarly, $x_{i+1}'$ is within distance $\e 2^s$ of $x_{i+1}$, where distance is measured with respect to the internal separator of $R_{i+1}$. 
    While this separator is not in $R_i$, it is in $R_i^{\arcto}$, thus $\dist_{R_i^{\arcto}}(x_{i+1}, x_{i+1}') \le \e 2^s$.
    Finally, observe that the subpath $\tilde P[x'_i:x'_{i+1}]$ is in $R_i^{\arcto}$ by construction, 
    where $\tilde P[x'_i:x'_{i+1}]$ denotes the subpath of $\tilde P$ beginning at $x'_i$ and ending at $x'_{i+1}$. By triangle inequality, we have 
    \[\dist_{R_i^{\arcto}}(x_{i}, x_{i+1}) \le \norm{\tilde P[x'_{i}:x'_{i+1}]} + O(\e 2^s).\]
    We conclude that $\sum_{i=1}^{k - 1} \dist_{R_i^{\arcto}}(x_i, x_{i+1}) \le \norm{\tilde P} + O(\log n)\cdot \e \cdot \norm{\tilde P}$. Again by triangle inequality and definition of $x_1$, we have that $\norm{\tilde P} \le (1+O(\e))\cdot \dist_{R}(a, p)$. Thus, 
    \[
    \sum_{i=1}^{k - 1} \dist_R(x_i, x_{i+1}) \le (1+O(\log n)\cdot\e)\cdot \dist_{R}(a, p).
    \]
    To conclude, we concatenate the exact canonical sequence from $a$ to $x_1$ with the canonical sequence $(x_1, \ldots, x_k)$. This provides a $(1+O(\log n)\cdot\e)$-canonical sequence from $a$ to $p$, which consists of $k = O(\log n)$ canonical pairs from $\operatorname{RelPairs}(a)$ as desired.
\end{proof}

\section{Proof of Lemma~\ref{lem:impl}: Proxy pairs}
\label{S:proxy}

The goal of this section is to prove \Cref{lem:impl}. Before we begin, we prove one claim about relevant pairs which will be useful in our algorithm; it follows directly from \Cref{clm:rel-helper}.

\begin{claim}
\label{clm:pairs-along-separator}
    Let $R$ be a region with internal separator $S$. Let $a$ and $b$ be two vertices on $S$. There is an (exact) 1-canonical sequence $(x_1, \ldots, x_\ell)$ between $a$ and $b$ with respect to $R$, such that (1) $\ell = O(\log \spread)$, (2) for every $i$, vertex $x_i$ lies on $S$, and (3) for every $i$, we have that $\canon{x_i}{x_{i+1}}$ is a canonical pair\footnote{Recall that the definition of canonical sequence only guarantees that either $\canon{x_i}{x_{i+1}}$ or $\canon{x_{i+1}}{x_{i}}$ is a canonical pair. We include the statement (3) mainly for notational convenience later.}.
\end{claim}
\begin{proof}
    This follows from \Cref{clm:rel-helper}. Define \EMPH{$i$} to be the scale such that $\dist_{R}(a,b) = \dist_S(a,b) \in [2^{i-1}, 2^i)$. 
    Clearly there is some $(i, R)$-portal {$p$} that lies on the path $S$ between $a$ and $b$. This portal $p$ is in both the relevant vertex sets $X_a$ and $X_b$ for $a$ and $b$ respectively; this is because $p$ lies on the shortest path $S$ between $a$ and $b$, so $\dist_{R}(a, p) \le \dist_S(a,b) \le 2^i$ and likewise $\dist_{R}(b, p) \le 2^i$. \Cref{clm:rel-helper} now guarantees the existence of an canonical sequence from $a$ to $p$, and from $b$ to $p$, each consisting of $O(\log \spread)$ vertices that lie on $S$. Concatenate these two canonical sequences to obtain an exact canonical sequence $(x_1, \ldots, x_\ell)$ between $a$ and $b$. Now, because every $x_i$ lies on $S$, both $\canon{x_i}{x_{i+1}}$ and $\canon{x_{i+1}}{x_{i}}$ are canonical pairs of $R$. This proves the claim.
\end{proof}

We are now ready to describe the procedure $\textsc{FindProxyPairs}(\canon a b)$, in Figure~\ref{fig:proxy}. It takes as input a canonical pair \EMPH{$\canon a b$}, and returns (1) a set of canonical pairs, which we denote \EMPH{$\proxy(\canon a b)$}, and (2) an approximate shortest path between $a$ and $b$, which we denote \EMPH{$\hat P(\canon a b)$}.
%
\begin{figure}[t]
\centering
\small
\begin{algorithm}
\textul{\textsc{FindProxyPairs}$(\canon a b)$}:\+
\\  $k \gets 0$
\\  $R_0 \gets$ region of the canonical pair $\canon a b$
\\  $\hat P_0 \gets \textsc{DetourPath}(\canon a b)$
\\  $\mathrm{Proxy}_0 \gets$ the set $\set{\canon a b}$
\\ ~
\\  while region $R_k$ is not the root of the separator hierarchy:\+
\\      $R_{k+1} \gets$ parent of $R_k$
\\      $S_{k+1} \gets$ internal separator of $R_{k+1}$
\\      $s \gets$ first vertex along $\hat P_k$ where $s \in S_{k+1}$
\\      $t \gets$ last vertex along $\hat P_k$ where $t \in S_{k+1}$
\\      $(x_1, x_2, \ldots, x_\ell) \gets$ a $1$-canonical sequence between $s$ and $t$ w.r.t. $R_{k+1}$, from Claim~\ref{clm:pairs-along-separator}
\\  \Comment{Note that $x_1 = s$ and $x_\ell = t$}
\\      $\hat P_{k+1} \gets \hat P_k[a:s] \circ \textsc{DetourPath}(\canon {x_1} {x_2}) \circ \ldots \circ \textsc{DetourPath}(\canon {x_{\ell-1}} {x_{\ell}}) \circ \hat P_k[t:b]$
\\      $\mathrm{Proxy}_{k+1} \gets \mathrm{Proxy}_{k} \cup \set{\canon {x_1} {x_2}, \ldots, \canon {x_{\ell-1}} {x_\ell}}$
\\      $k \gets k + 1$\-
\\ return $(\mathrm{Proxy}_k, \hat P_k)$
\end{algorithm}
\caption{Procedure $\textsc{FindProxyPairs}(\canon a b)$}
\label{fig:proxy}
\end{figure}
%


\begin{observation}
    $\mathrm{Proxy}(\canon a b)$ contains $O(\log n \cdot \log \spread)$ canonical pairs.
\end{observation}

\begin{proof}
    The separator hierarchy has height $O(\log n)$, so there are $O(\log n)$ iterations in $\textsc{FindProxyPairs}(\canon a b)$. Each iteration adds $O(\log \spread)$ canonical pairs to the proxy set, by Claim~\ref{clm:pairs-along-separator}.
\end{proof}


We now show that the union of paths in $\safe(\canon {a'} {b'})$ over all $(\canon{a'}{b'}) \in \mathrm{Proxy}(\canon a b)$ contains an approximate shortest path between $a$ and $b$.
\begin{lemma}
    The path $\hat P(\canon a b)$ is contained in $\bigcup_{(\canon {a'}{b'}) \in \mathrm{Proxy}_k} \safe(\canon {a'} {b'})$.
\end{lemma}

\begin{proof}
    For any $k\ge 0$, recall that $\textsc{FindProxyPairs}(\canon a b)$ iteratively defines the region \EMPH{$R_k$}, its internal separator \EMPH{$S_k$}, the path \EMPH{$\hat P_k$}, and the set \EMPH{$\proxy_k$}.
    We show by induction on $k$:
    \begin{quote}
        The path $\hat P_k$ is contained in the union of $\bigcup_{(\canon {a'}{b'}) \in \mathrm{Proxy}_k} \safe(\canon {a'}{b'})$ and all external separators of $R_k$.
    \end{quote}
    This claim suffices to prove the lemma: when the $\textsc{FindProxyPairs}$ procedure terminates at some iteration $k'$ and returns $\hat P(\canon a b) =P_{k'}$, the region $R_{k'}$ has no proper ancestors and thus has no external separators.

    \medskip \noindent
    In the \textbf{base case}, we have $k = 0$, meaning $R_0$ is the region of the canonical pair $\canon a b$, path $\hat P_0 = \textsc{DetourPath}(\canon a b)$ and $\mathrm{Proxy}_0 = \set{\canon a b}$. By Lemma~\ref{lem:detour-subpath}, $\hat P_0$ is contained in the union of $R_0$ and all external separators of $R_k$.

    \medskip \noindent
    In the \textbf{inductive case}, we assume the claim is true for path $\hat P_{k}$ and prove it is true for $\hat P_{k+1}$.
    Recall that by definition, $\hat P_{k + 1}$ can be written as the union of subpaths
    \[
    \hat P_{k+1} = \hat P_{k}[a:s] \circ \textsc{DetourPath}(\canon {x_1} {x_2}) \circ \ldots \circ \textsc{DetourPath}(\canon {x_{\ell-1}} {x_\ell}) \circ \hat P_k[t:b].
    \]
    For each of these subpaths, we prove that it is contained in the union of $\bigcup_{(\canon{a'}{b'}\in \proxy_k)}\safe(\canon{a'}{b'})$ and all external separators of $R_{k+1}$.
    We first prove the claim for the subpath $\textsc{DetourPath}(\canon {x_i} {x_{i+1}})$, for any $i$.
    Observe that the canonical pair $\canon {x_i} {x_{i+1}}$ is in region $R_{k+1}$ by Claim~\ref{clm:pairs-along-separator}(2); 
    thus, the [subpath decomposition] property (Lemma~\ref{lem:detour-subpath}) of $\textsc{DetourPath}$ guarantees that $\textsc{DetourPath}(\canon {x_i} {x_{i+1}})$ is contained in the union of $\safe(\canon {x_i} {x_{i+1}})$ and all external separators of $R_{k+1}$. The fact that $\canon {x_i} {x_{i+1}}$ is added to $ \mathrm{Proxy}_{k+1}$ proves the claim.
    Now we consider the subpaths $\hat P_{k}[a:s]$ and $\hat P_{k}[t:b]$. By induction, $\hat P_{k}[a:s]$ and $\hat P_{k}[t:b]$ are contained in the union of $\bigcup_{(\canon {a'} {b'}) \in \mathrm{Proxy}_k} \safe(\canon {a'}{b'})$ and all external separators of $R_k$. But by choice of $s$ (resp.\ $t$), no edge of $\hat P_{k}[a:s]$ (resp.\ $\hat P_{k}[t:b]$) is contained in the separator $S_{k+1}$. Every external separator of $R_k$ other than $S_{k+1}$ is also an external separator of $R_{k+1}$ (as $R_{k+1}$ is the parent of $R_k$), and every canonical pair $(\canon{a'} {b'}) \in \mathrm{Proxy}_k$ is also in $\mathrm{Proxy}_{k+1}$. Thus, every edge of $\hat P_{k}[a:s]$ (resp.\ $\hat P_{k}[t:b]$) contained in the union of $\bigcup_{(\canon{a'}{b'}) \in \mathrm{Proxy}_{k+1}} \safe(\canon{a'}{b'})$ and all external separators of $R_{k+1}$.
\end{proof}

It remains to prove a bound on the length of $\hat P_k$. The structure of the argument is essentially a variant of the proof of the [stretch] property of \textsc{DetourPath} (Lemma~\ref{lem:safe-stretch}), and crucially depends on Lemma~\ref{lem:safe-stretch} as a base case. We begin by giving more general notation to the concepts of $\domain(\cdot)$ and $\replace {\cdot}{\cdot}$ from Lemma~\ref{lem:safe-stretch}; the differences are underlined.
For any path $P$ and subpath $X = P[x_1:x_2]$ of $P$,
we define the \EMPH{domain} of $X$ \textul{with respect to region $R$}, denoted \EMPH{$\domain_R(X)$}, to be the lowest region $R'$ in the separator hierarchy such that (1) $R'$ contains both \emph{endpoints} of $X$, and (2) $R'$ is a (not necessarily proper) ancestor \textul{of $R$}.
For any set of edge-disjoint subpaths $\cX$ of $P$, we define the \EMPH{domain replacement} of $P$ for $\cX$ \textul{with respect to $R$}, denoted \EMPH{$\domreplace {R} P {\cX}$}, to be the path obtained by starting with $P$ and replacing every subpath $X \in \cX$ with a shortest path \textul{in $\domain_{R}(X)$} between the endpoints of $X$. 
We need this new notation because, in this section, we will be working with canonical pairs $\canon {a'}{b'}$ that are all associated with different regions.
In Section~\ref{S:safe}, the region $R$ was implicit, as we only analyzed $\textsc{DetourPath}$ applied to some fixed canonical pair in some fixed region.
With our new notation, Lemma~\ref{lem:safe-stretch} and Corollary~\ref{cor:safe-stretch-simple} can be restated as follows:

\begin{corollary}[Restatement of Lemma~\ref{lem:safe-stretch} and Corollary~\ref{cor:safe-stretch-simple}]
\label{cor:general-stretch-detour}
There is a value $\beta = O(\log n \cdot \log \spread)$ such that the following holds:
Let $\canon {a} {b}$ be a canonical pair with region $R$ (and canonical region $R^{\arcto}$), and let $\pi$ be the path returned by $\textsc{DetourPath}(\canon a b)$. Then for any set $\cX$ of subpaths of $P$,
\[ 
\norm{\domreplace{R}{P}{\cX}} \le (1+\beta \cdot \e) \cdot \dist_{R^{\arcto}}(a,b).
\]
\end{corollary}    

We will use Corollary~\ref{cor:general-stretch-detour} as a base case to prove a stretch bound on the paths $\hat P_k$ constructed throughout \textsc{FindProxyPairs}.

\begin{lemma}
Let $\beta = O(\log n \cdot \log \spread)$ be the value from Corollary~\ref{cor:general-stretch-detour}. 
Let $k\ge 0$.
For any set $\cX$ of subpaths of $\hat P_k$,
\[
\norm{\domreplace{R_k}{\hat P_k}{\cX}} \le (1+ 2(k+1) \beta\e)\cdot \dist_{R_0^{\arcto}}(a,b)
\]
where $R_0^{\arcto}$ denotes the canonical region of $\canon a b$.
\end{lemma}

\begin{proof}
The proof is by induction on $k$. In the \textbf{base case} of $k = 0$, path $\hat P_0$ is simply $\textsc{DetourPath}(\canon a b)$, and $R_0$ is the region of the canonical pair $\canon a b$. The claim follows from Corollary~\ref{cor:general-stretch-detour}.

\medskip \noindent
In the \textbf{inductive case}, we have $k > 0$ and assume that the claim holds for $\hat P_{k-1}$. 
By construction, $\hat P_k$ can be written as
\[
\hat P_k = \hat P_{k-1}[a:x_1] \circ \textsc{DetourPath}(\canon {x_1} {x_2}) \circ \ldots \circ \textsc{DetourPath}(\canon {x_{\ell-1}} {x_\ell}) \circ \hat P_{k-1}[x_\ell:b].
\]
where $(x_1, \ldots, x_\ell)$ is the (exact) $1$-canonical sequence between $x_1 = s$ and $x_\ell = t$, with respect to $R_k$, which was constructed in the $k$th iteration of \textsc{FindProxyPairs} by \Cref{clm:pairs-along-separator}.

Notice that every vertex $x_1, \ldots x_\ell$ is in region $S_k$ by Claim~\ref{clm:pairs-along-separator}(2), and thus in region $R_k$.
We may assume WLOG that every subpath in $\cX$ is fully contained in either $\hat P_k[a:x_1]$, $\hat P_k[x_1:x_2]$, $\ldots$, $\hat P_k[x_{\ell-1}:x_\ell]$, or $\hat P_k[x_\ell:b]$. 
Indeed, if some subpath $X = \hat P_k[u:v] \in \cX$ was not contained in one of the subpaths described above, then there is some vertex $x_i$ that lies strictly between $u$ and $v$ along $P$; we could remove $P_k[u:v]$ from $\cX$ and instead add in the two subpaths $X_1 = P_k[u:x_i]$ and $X_2 = P_k[x_i:v]$. 
As $x_i$ is in $R_k$, the subgraph $\domain_{R_k}(X)$ contains $\domain_{R_k}(X_1)$ and $\domain_{R_k}(X_2)$, triangle inequality implies that swapping out $\set{X_1, X_2}$ instead of $X$ can only increase the length of $\domreplace{R_k}{\hat P_k}{\cX}$. 
Our goal is to upper bound this length, so we may make the WLOG assumption.

For any two vertices $u$ and $v$, let \EMPH{$\cX_{u:v}$} denote the set of subpaths in $\cX$ that are entirely contained in $\hat P_k[u:v]$, and let \EMPH{$\tilde{P}[{u:v}]$} denote $\domreplace{R_k}{\hat P_k[u:v]}{\cX_{u:v}}$. By our assumption from the previous paragraph, we have
$\domreplace{R_k}{\hat P_k}{\cX} = \tilde P[{a:x_1}] \circ \tilde P[{x_1:x_2}] \circ \ldots \circ \tilde P[{x_{\ell-1}:x_\ell}] \circ \tilde P[{x_\ell:b}]$, and so 
\begin{equation}
\label{eq:proxy-decomposition}
    \norm{\domreplace{R_k}{\hat P_k}{\cX}} = \norm{\tilde P[{a:x_1}]} + \sum_{i= 1}^{\ell - 1} \, \norm{\tilde P[{x_i:x_{i+1}}]}
    + \norm{\tilde P[{x_\ell:b}]}.
\end{equation}
First observe that the induction hypothesis implies that
\begin{equation}
\label{eq:proxy-induction}
    \norm{\tilde P[{a:x_1}]} + \dist_{R_k}(x_1, x_\ell) + \norm{\tilde P[{x_\ell:b}]} \le (1 + 2 \beta k \e) \cdot \dist_{R_0^{\arcto}}(a,b).
\end{equation}
Indeed, define $\EMPH{$\cX'$} \coloneqq \cX_{a:x_1} \cup \set{\hat P_{k-1}[x_1:x_\ell]} \cup \cX_{x_\ell:b}$.
Observe that (by choice of $x_1$ and $x_\ell$) $\cX'$ is a set of subpaths of $\hat P_{k-1}$. 
Thus $\domreplace{R_{k-1}}{\hat P_{k-1}}{\cX'}$ is well-defined.
Observe that $\domain_{R_{k-1}}(\hat P_{k-1}[x_1:x_\ell]) = R_k$, as $x_1$ and $x_\ell$ lie on the internal separator of $R_k$, and $R_k$ is an ancestor of $R_{k-1}$. 
Further, for any subpath $X \in \cX$, we have that $\domain_{R_{k-1}}(X)$ is contained in $\domain_{R_{k}}(X)$; in particular, any shortest path in $\domain_{R_{k-1}}(X)$ is no longer than the corresponding shortest path in $\domain_{R_{k}}(X)$. Thus $\norm{\domreplace{R_{k-1}}{\hat P_{k-1}}{\cX'}} \ge \norm{\tilde P[{a:x_1}]} + \dist_{R_k}(x_1, x_\ell) + \norm{\tilde P[{x_\ell:b}]}$. Applying induction hypothesis gives \eqref{eq:proxy-induction}.

\medskip 
\noindent Finally, we show that
\begin{equation}
\label{eq:proxy-canonical-stretch}
    \sum_{i=1}^{\ell-1} \,  \norm{\tilde P[{x_i: x_{i+1}}]} \le \dist_{R_k^{\arcto}}(x_1, x_\ell) + 2 \beta \e \cdot \dist_{R_0^{\arcto}}(a,b).
\end{equation}
Indeed, the fact that $(x_1, \ldots, x_\ell)$ is a $1$-canonical sequence w.r.t. $R_k$ implies that $\sum_{i=1}^{\ell-1} \dist_{R_k}(x_i, x_{i+1}) = \dist_{R_k}(x_1, x_\ell)$.\footnote{The definition of canonical sequence actually bounds $\dist_{R_k^{\arcto}}(x_i, x_{i+1})$, where $R_k^{\arcto}$ is the canonical subgraph of $\canon{x_i}{x_{i+1}}$. But because $x_i$ and $x_{i+1}$ are both on the internal separator $S_k$ (by \Cref{clm:pairs-along-separator}(2)), we have that $R_k = R_k^{\arcto}$.}
Corollary~\ref{cor:general-stretch-detour} implies that each $\tilde P[{x_i:x_{i+1}}]$ has length at most $(1+\beta \e) \cdot \dist_{R_k^{\arcto}}(x_i, x_{i+1})$, and so
\[
\sum_{i=1}^{\ell - 1} \, \norm{\tilde P[{x_i:x_{i+1}}]} \le (1 + \beta \e) \cdot \dist_{R_k^{\arcto}}(x_1, x_\ell).
\]
To prove Equation~\ref{eq:proxy-canonical-stretch}, it remains to show that $\dist_{R_k^{\arcto}}(x_1, x_\ell) \le 2\dist_{R_0^{\arcto}}(a,b)$. Indeed, Equation~\eqref{eq:proxy-induction} implies that $\dist_{R_k}(x_1, x_\ell) \le (1+2\beta k \cdot \e) \cdot \dist_{R_0^{\arcto}}(a,b) \le 2 \cdot \dist_{R_0^{\arcto}}(a,b)$ provided that $\e \le \frac{1}{2 \cdot\beta k}$. The condition on $\e$ is satisfied because $\e = \frac{\e_0}{\Theta(\log^3 n \cdot \log \spread)}$ for some $\e_0 < 1$, and $\beta = O(\log n \log \spread)$ and $k = O(\log n)$.
This proves Equation~(\ref{eq:proxy-canonical-stretch}).
Equations~\eqref{eq:proxy-decomposition}, \eqref{eq:proxy-induction}, \eqref{eq:proxy-canonical-stretch}, and the fact that $\dist_{R_k^{\arcto}}(x_1, x_\ell) \le \dist_{R_k}(x_1, x_\ell)$ together prove the inductive claim.
\end{proof}

\begin{corollary}
    The path $\hat P(\canon a b)$ has length at most $(1+O(\log^2 n \cdot \log \spread)\cdot \e)\cdot \dist_{R_0^{\arcto}}(a,b)$.
\end{corollary}
This completes the proof of \Cref{lem:impl}.

\section{Fast construction}
\label{S:fast}
In this section we give a near-linear-time construction for our DAM, proving \Cref{thm:fast-full-statement}. We do this with a general bootstrapping argument, similar to \cite{ckt-lpg-2022}.
The key tool we use is $r$-division. An \EMPH{$(r, s)$-division} of a graph $G$ is a partition of the edges of $G$ into $O(n/r)$ subgraphs called regions, such that each region $R$ (1) has at most $r$ vertices and (2) shares at most $s$ \EMPH{boundary vertices} (denotes $\partial R)$ with other regions. The notion of $(r, O(\sqrt r)$-division first appeared in \cite{fre-faspp-1987} (where it was simply called $r$-division) in the context of planar graphs; the more general notation of $(r, s)$-division was introduced by \cite{hkrs-fsapg-1997}. For minor-free graphs, \cite{HR24} show
that an $(r, O(r^{2/3} \log^{1/3} r))$-division can be constructed in $O(n)$ time. Throughout this section, we will use the term \EMPH{$r$-division} to mean the $(r, O(r^{2/3} \log^{1/3}r))$-division of \cite{HR24}.

\thmFastFull*

\begin{proof}
By \Cref{thm:dam-scaled} there is a $\poly(n, \log \Phi, \e^{-1})$-time algorithm to construct a $(1+\e)$-DAM for $(G, T)$, for any minor-free graph $G$ and terminal set $T$.  Let $\EMPH{$\kappa(n, \Phi, \e)$} = \Theta(\e^{-7} (\log n)^{30} (\log n\Phi)^{13})$; that is, the DAM of \Cref{thm:dam-scaled} is guaranteed to have size at most $\kappa(n, \Phi, \e) \cdot |T|$.

Given a graph $G$ and a terminal set $T$, we construct a $(1+\e)$-DAM for $(G, T)$ with an iterative procedure, which iteratively constructs DAMs $G_i$ of decreasing size.
Let $\e' \coloneqq \frac{\e}{\Theta(\log n)}$, and abbreviate $\EMPH{$\kappa$} \coloneqq \kappa(n, \Phi, \e')$.
To begin, we set $\EMPH{$G_0$} \coloneqq G$, observing that $G_0$ is a trivial DAM for $(G, T)$ with stretch $1$ and size $n$; and initialize the iteration counter to $i \gets 0$. At each iteration $i$, we consider graph \EMPH{$G_i$}, which we inductively guarantee is a DAM for $(G, T)$.  If $G_i$ has fewer than $4 \kappa \cdot |T|$ vertices, then $G_i$ itself is a small-size DAM and we are done. Otherwise, let $\EMPH{$r$} \coloneqq \kappa^4$. Construct an $r$-division for $G_i$, and let $\mathcal R$ denote the set of regions. For each region $R \in \mathcal R$, we define a terminal set $T_R = \bdry R \cup (R \cap T)$ to be the set of boundary vertices of $R$ plus the terminals inside $R$, and replace $R$ with a $(1+\e')$-DAM for $(R, T_R)$. This produces a new graph \EMPH{$G_{i+1}$}; observe that $G_{i+1}$ is a minor of $G_i$ that contains all terminals in $T$ as vertices. Increment $i \gets i+1$ and repeat.

\smallskip \noindent \textbf{Runtime.} Consider each iteration $i$. Let $n_i$ denote the number of vertices of $G_i$. The $r$-division can be found in $O(n_i)$ time, and it takes $\poly (r, \log \Phi, (\e')^{-1}) = \poly (\log n, \log \Phi, \e^{-1})$ time to construct a DAM for each of the $O(n_i/r)$ pieces; thus the entire replacement process takes at most $O(n_i \cdot \poly (\log n, \log \Phi, \e^{-1}))$ time. The new graph $G_{i+1}$ has size at most
\begin{align*}
    \sum_{R \in \mathcal R} \kappa \cdot |T_R|
    &\le \sum_{R \in \mathcal R} \kappa \cdot  \left(|T \cap R| + O(r^{2/3} \log^{1/3} r) \right)\\
    &\le \kappa \cdot \left(|T| + O\left(\frac{n_i}{r}\right) \cdot O(r^{2/3} \log^{1/3} r)\right)\\
    &\le  \kappa \cdot |T| + O\left(\frac{\kappa \cdot \log^{1/3} r }{r^{1/3}}\right)\cdot n_i\\
    &\le \frac{n_i}{4} + \frac{n_i}{4} \\
    &\le \frac{n_i}{2}
\end{align*}
where the second-to-last inequality holds because (1) we assume $n_i \ge 4\kappa \cdot |T|$ and (2) because we chose $r > \kappa^4$ (and so $O(\frac{\kappa \cdot \log^{1/3} r}{r^{1/3}}) \le \frac{1}{4}$). Thus, the size of the graph shrinks by at least a constant factor in each iteration. The recursion terminates after $O(\log n)$ levels, and the overall runtime is $O(n \cdot \poly (r, \log \Phi, \e^{-1})) = O(n \cdot \poly (\log n, \log \Phi, \e^{-1}))$.

\smallskip \noindent \textbf{Stretch.} Notice that distances between terminals in $G_{i+1}$ are distorted by at most a $(1+\e')$ factor compared to $G_i$. Indeed, fix two terminals $t_1$ and $t_2$, and let $\pi$ be a shortest path between $t_1$ and $t_2$ in $G$. The path $\pi$ can be partitioned into subpaths each of which is entirely contained in some region $R$ of the $r$-division, and the endpoints of each subpath are either terminals or boundary vertices. After replacing each region with a DAM, the length of each subpath is distorted by a $(1+\e')$, leading to overall distortion $(1+\e')$ in $G_{i+1}$. Now, over all $O(\log n)$ levels of recursion, the distortion accumulates to $(1+\e')^{O(\log n)} \le (1+O(\log n)\e') \le (1 + \e)$.
\end{proof}


\small
\bibliographystyle{alphaurl}
\bibliography{dam}

\end{document}